\pdfoutput=1
\documentclass[fleqn,usenatbib]{mnras}

\usepackage{newtxtext,newtxmath}
\usepackage{array}
\usepackage[dvipsnames]{xcolor}
\newcolumntype{P}[1]{>{\centering\arraybackslash}p{#1}}
\usepackage{lineno}
\usepackage{dirtytalk}

\usepackage[T1]{fontenc}
\usepackage{ae,aecompl}
 \usepackage{transparent}

\usepackage{graphicx}	
\usepackage{amsmath}	
\usepackage{float}

\title[Shock-cloud interactions with TI and gravity]{Interactions of a shock with a molecular cloud at various stages of its evolution due to thermal instability and gravity}

\author[M. M. Kupilas et al.]{
M. M. Kupilas,$^{1}$\thanks{Email: py12mk@leeds.ac.uk}
C. J. Wareing,$^{1}$
J. M. Pittard,$^{1}$
S. A. E. G. Falle$^{2}$
\\
$^{1}$School of Physics and Astronomy, University of Leeds, Leeds LS2 9JT, UK\\
$^{2}$Department of Applied Mathematics, University of Leeds, Leeds LS2 9JT, UK\\
}

\date{Accepted 2020 November 30. Received 2020 November 18; in original form 2020 October 8}

\pubyear{2020}

\begin{document}
\label{firstpage}
\pagerange{\pageref{firstpage}--\pageref{lastpage}}
\maketitle

\begin{abstract}
\noindent\textcolor{blue}{This is a pre-copyedited, author-produced PDF of an article accepted for publication in \textit{Monthly Notices of the Royal Astronomy Society} following peer review.}

\vspace{2mm}
Using the adaptive mesh refinement code MG, we perform hydrodynamic simulations of the interaction of a shock with a molecular cloud evolving due to thermal instability and gravity. To explore the relative importance of these processes, three case studies are presented. The first follows the formation of a molecular cloud out of an initially quiescent atomic medium due to the effects of thermal instability and gravity. The second case introduces a shock whilst the cloud is still in the warm atomic phase, and the third scenario introduces a shock once the molecular cloud has formed. The shocks accelerate the global collapse of the clouds with both experiencing local gravitational collapse prior to this. When the cloud is still atomic, the evolution is shock dominated and structures form due to dynamical instabilities within a radiatively cooled shell. While the transmitted shock can potentially trigger the thermal instability, this is prevented as material is shocked multiple times on the order of a cloud crushing time-scale. When the cloud is molecular, the post-shock flow is directed via the pre-existing structure through low-density regions in the inter-clump medium. The clumps are accelerated and deformed as the flow induces clump-clump collisions and mergers that collapse under gravity. For a limited period, both shocked cases show a mixture of Kolmogorov and Burgers turbulence-like velocity and logarithmic density power spectra, and strongly varying density spectra. The clouds presented in this work provide realistic conditions that will be used in future feedback studies.

\end{abstract}




\section{Introduction}

The interstellar medium (ISM) is observed to exist in multiple phases, ranging from a hot ionized plasma (\textit{T}\,$\gtrsim$\,10$^6$\,K, \textit{n}\,$\lesssim$\,0.01\,cm$^{-3}$) to a cold molecular gas (\textit{T}\,$\sim$\,10\,--\,20\,K, \textit{n}\,$\gtrsim$\,100\,cm$^{-3}$) \citep{mckee1977theory,wolfire1995neutral,wolfire2003neutral}. 
Stars form in the densest regions, which themselves are part of giant molecular clouds (GMCs). GMCs are clumpy and filamentary \citep{andre2014filamentary} and have structures long known to be characterised by a turbulent velocity field \citep{larson1981turbulence}.

This dynamical state of GMCs must be partly driven and maintained by physical processes that form them and their substructures. On galactic scales, spiral-arm compression \citep[][]{dobbs2008ism,dobbs2008simulations}, magneto-rotational instability \citep{balbus1991powerful,tamburro2009driving} and galactic shear drive solenoidal flows whose energies cascade down to smaller scales contributing to the supersonic motions observed. Accretion due to gravitationally driven flows and subsequent fragmentation can drive compressive supersonic motions on galactic scales down to scales of molecular clouds and pre-stellar cores \citep{hoyle1953fragmentation,field2008model,klessen2010accretion,federrath2011new,van2014core}. On molecular cloud scales, collisions with other clouds can also drive their internal turbulence \citep{wu2018giant}, and trigger the star formation process \citep[][]{tan2000star,wu2015gmc,wu2017gmc2,wu2017gmc3}. Once stars form, GMCs are subsequently affected from within by the feedback from the most massive stars in the form of winds, radiation and supernovae \citep[e.g.][]{rogers2013feedback,wareing2016magnetohydrodynamic,wareing2017hydrodynamic,wareing2018new}. These processes affect the state of their parent cloud, and material escaping to the wider ISM regulates its subsequent state thus affecting future generations of stars. The currently developing consensus to understand this cycle is the turbulence-regulated paradigm of star formation \citep{mac2004control,elmegreen2004interstellar,federrath2012star} where the dominant driving mechanisms at different stages are currently under debate.

Adding to the complexity of the problem, the strong density inhomogeneities observed in the ISM can also be attributed to thermal phase transitions due to inherent instabilities resulting from the balance of heating and cooling processes \citep{parker1953instability,field1965thermal,field1969cosmic,wolfire1995neutral,wolfire2003neutral}. This thermal instability (TI) can develop in association with other processes such as externally driven turbulence \citep[e.g.][]{saury2014structure}, spiral-arm shocks \citep[e.g.][]{yang2012thermal,kim2008galactic,kim2010galactic}, gravity and magnetic fields \citep[][hereafter WPFVL16]{wareing2016}. In fact, the TI in isolation has been demonstrated to be an effective mechanism for converting warm diffuse gas into cold and dense material, generating conditions for star formation and driving large scale flows that result in Kolmogorov power spectra \citep*[][hereafter WFP19,WPF20]{wareing2019,Wareing2020}.

To study the TI in isolation, WPFVL16 performed 3D simulations of a quiescent 17\,000 M$_\odot$ cloud seeded with $\pm$\,10 per cent density perturbations around an equilibrium state of \textit{n}\,=\,1.1\,cm$^{-3}$ in the warm unstable phase. The simulations performed involved successive increases in complexity by including gravity and magnetic fields of different strengths, neglecting any additional mechanisms such as driven turbulence, converging flows or feedback. 
They found that after $\sim$\,20\,Myrs of evolution the growth of the density perturbations formed clumps (and in the magnetic field cases, filaments) with properties that connected well with observations of molecular clouds. A higher resolution hydrodynamic (HD) study by WFP19 of a more massive cloud explored the interplay between gravity and the TI and found that realistic clump masses, sizes, velocity dispersions and power spectra could be achieved without resorting to additional driving mechanisms. Additionally, the clumps were connected by 0.3\,--\,0.5\,pc width filaments that continuously fed material to the clumps and as the original cloud had 8$\,\times$ the mass of WPFVL16, the clumps were able to gather enough mass to collapse under gravity and conclude the star formation process. Most recently, a higher resolution study by WPF20 including TI, magnetic fields and gravity now also provides possible explanations for the origins of features such as striations \citep{goldsmith2008large}, hour glass magnetic fields \citep{pattle2017jcmt} and the integral shaped filament (ISF) in the Orion A molecular cloud \citep{stutz2016slingshot,stutz2018integral}.

The studies of WPFVL16/WFP19 effectively demonstrated the importance of the TI in the process of molecular cloud and star formation. However, as the models were highly idealised it is now appropriate to incrementally introduce extra dynamical ingredients. For example, as the time-scales of evolution of the models in WPFVL16 are long and the clouds were evolved for a free-fall time of $\sim$\,45\,Myrs, it is  likely that in reality a cloud like this would experience one or more shocks. 

The interaction of shocks with clouds is a ubiquitous problem already studied by many authors so the basic physics of the interaction is well understood. Several works have explored various scenarios such as adiabatic interactions \citep[e.g.][]{klein1994hydrodynamic}, effects due to radiative cooling \citep[e.g.][]{fragile2004radiative,yirak2010self}, magnetic fields \citep[e.g.][]{maclow1994ApJ...433..757M, fragile2005magnetohydrodynamic}, sub-grid turbulence \citep[e.g.][]{pittard2009turbulent,pittard2010turbulent,pittard2016turbulent} and different cloud profiles \citep[e.g.][]{nakamura2006hydrodynamic} and shapes \citep[e.g.][]{pittard2016kathryn,goldsmith2016interaction,goldsmith2020isothermal}. Often, these works mainly focused on the effects of different physics on the particulars of the cloud destruction. However, the compression due to a shock can also be an effective mechanism for generating cold dense clouds out of warm diffuse gas, triggering the thermal instability and in the process generating the conditions for star formation \citep[e.g.][]{inoue2009two,aota2013thermal,inoue2015thermal}.

For example, \citet{van2007shock,van2010shock} explored in both 2D and 3D the effects of a shock interacting with diffuse atomic clouds initially in a warm stable state with a density of \textit{n}\,=\,0.45\,cm$^{-3}$. They included an adapted cooling function and magnetic fields, and found that depending on the magnetic field orientation relative to the shock normal, magnetically dominated clouds formed with properties that resembled those of molecular clouds and low-density HI clouds. Since the effects of heating and cooling were also included, the transmitted fast mode shocks demonstrated an ability to trigger the TI. Similar results were found by \citet{heitsch2009effects} which affirmed the utility of shock-cloud interactions in creating conditions ideal for the formation of dense molecular clumps. However, as gravity was not included, and their resolution was too low to fully resolve the cooling length, it was not possible to elucidate the importance of the TI and witness the formation of clumps and cores.

In this paper we build on the work of WPFVL16 which included gravity and sufficient resolution to track the behaviour of the TI, by introducing a shock. We thus present a self consistent HD model of molecular cloud formation due to the TI, gravity and shocks. We track the evolution of a cloud from initially quiescent atomic gas, to the formation of clumps that eventually collapse under gravity, demonstrating the ability of the clouds to form stars. During their evolution we explore the relative importance of the physical processes driving their behaviour. In the next Section we present the numerical method and the models studied. In Section 3 we present the results and discussion, and Section 4 provides a summary and conclusion.

\section{Setup}
\subsection{Numerical method}

This work presents 3D hydrodynamic simulations of the interaction of a shock with a cloud that is evolving due to the thermal instability and gravity. All calculations were performed using the adaptive mesh refinement (AMR) code MG \citep{falle2005amr,hubber2013convergence}.

The AMR method initialises a computational domain with a hierarchy of grids \textit{G}$_0$ ... \textit{G}$_{N-1}$ where \textit{N} is the number of levels chosen for the simulation. Levels \textit{G}$_0$ and \textit{G}$_1$ are initialised by default and cover the whole domain. The fractional difference between solutions within a cell and its parent is used to control refinement, and on a cell-by-cell basis finer grids are created when this exceeds a given error tolerance. When the error in refined cells falls below the tolerance, the cells are removed. A cell on \textit{G}$_0$ has size $\Delta$\textit{x} and timestep $\Delta$\textit{t} and has size $\Delta$\textit{x}/2$^{N-1}$ and timestep $\Delta$\textit{t}/2$^{N-1}$ on the \textit{N}$^{\textup{th}}$ level in order to ensure the fluid step is synchronised at coarse and fine boundaries. Such a grid structure improves
the efficiency of the code by confining the fine grids to where they are needed.
The code solves the Euler equations of hydrodynamics and employs an upwind, conservative shock-capturing scheme. Every cell interface is treated as a Riemann problem and solved to obtain the conserved fluxes. These are used to integrate the solution in time according to a Godunov scheme \citep{godunov1959difference} and piece-wise linear cell interpolation and a predictor-corrector method make the code second order in space and time. Heating and cooling processes are included as an energy source calculated using a lookup table of cooling rates for a given temperature and self-gravity is computed using a full approximation multigrid. The potential is set to\,=\,0 on the boundaries, and free-flow boundary conditions are used for the fluid. For a detailed description of the numerical methods used and the precise cooling curve adopted please consult WPFVL16.

\subsection{Model}

\begin{figure}
\vspace*{-0.5cm}
\includegraphics[width=0.5\textwidth]{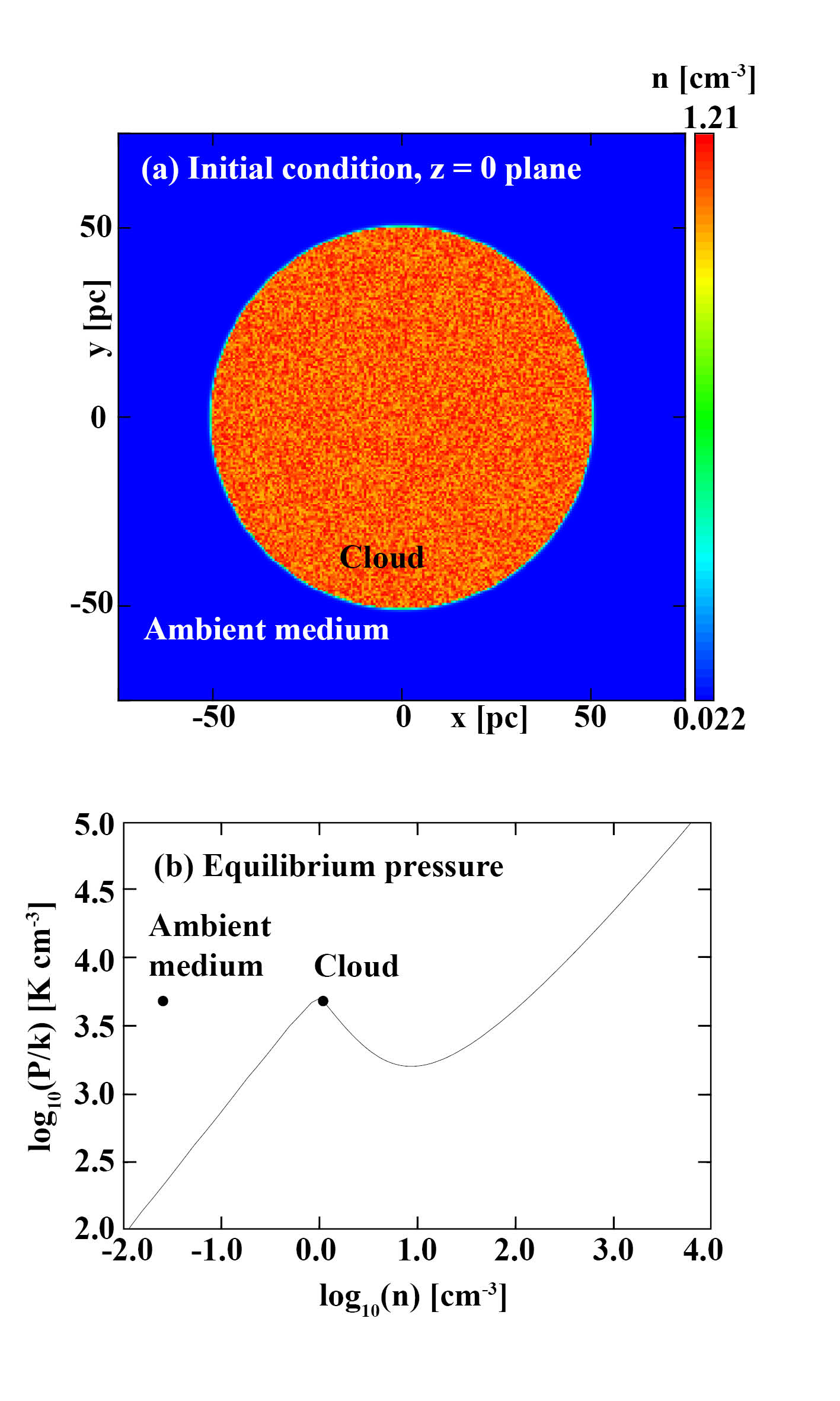}
\caption{(a): Initial condition showing density perturbations in a density slice through $\textit{z}$\,=\,0. (b): Equilibrium pressure against density for the warm, unstable and cold ISM. The initial condition of the cloud and ambient gas with $\chi$\,=\,50 are marked on the plot.}
\end{figure}

The model used for this work is similar to the set-up used in scenario 3 in WPFVL16. Namely, we initialise 17\,000 M$_\odot$ of diffuse material within a spherical cloud with a 50\,pc radius. The cloud lies at the centre of a uniform Cartesian domain of 300\,pc on each side with a numerical extent -3\,<\,\textit{xyz}\,<\,3. The cloud is seeded with random density variations of $\pm$\,10 per cent around \textit{n}\,=\,1.1\,cm$^{-3}$, with a pressure set according to the unstable equilibrium of heating and cooling at \textit{P}$_{\textup{eq}}$/\textit{k}\,=\,4800\,$\pm$\,300\,K\,cm$^{-3}$. We assume a mean particle mass of 2.0\,$\times$10$^{-24}$\,g. The initial condition at \textit{t}\,=\,0 and the dependence of the equilibrium pressure on the density with the cloud and ambient state are shown in Fig.\,1.

The cloud is embedded in a lower density medium with a density contrast $\chi$\,=\,50, where

\begin{equation}
    \chi=\frac{\textit{n}_{\textup{cl}}}{\textit{n}_{\textup{amb}}},
\end{equation}

\noindent thus setting the ambient medium with \textit{n}$_{\textup{amb}}$\,=\,0.022\,cm$^{-3}$. In order to keep the cloud confined, the surroundings are set with an equivalent pressure of \textit{P}/\textit{k}\,=\,4800\,K\,cm$^{-3}$, without the fluctuations, resulting in a temperature of 218\,000\,K. Gas in the ISM with these temperatures has a short cooling time of $\sim$\,0.2\,Myrs, which is significantly lower than the dynamical time-scale characterised by the free-fall time \textit{t}$_{\textup{ff}}$ defined as

\begin{equation}
\textit{t}_{\textup{ff}} = \sqrt{\frac{3\pi}{32G\rho}},
\end{equation}

\noindent which for \textit{n}\,=\,1.1\,cm$^{-3}$ is 45\,Myrs. For this reason the heating and cooling are switched off in the ambient medium to keep the cloud confined during its quiescent (shock-less) evolution. It is possible to confine the cloud in a lower density medium where \textit{t}$_{\textup{cool}}$\,$>$\,\textit{t}$_{\textup{ff}}$. However, as this raises the temperature and thus shortens the timestep, a choice of $\chi$\,=\,50 ensures the simulations are not computationally expensive. When the shock is introduced, the ambient medium is reset to \textit{n}\,=\,0.0022\,cm$^{-3}$ ($\chi$\,=\,500) and heating and cooling are switched on everywhere in the domain. Additionally, to prevent any numerical errors from developing due to sharp edges and large density contrasts, and since clouds in the ISM are unlikely to have sharp edges \citep[e.g. see discussion in][]{nakamura2006hydrodynamic}, we smooth out the cloud interface over $\sim$\,5 cells. Note that for our resolution, our clouds have $\sim$\,170 cells per radius, and so the smooth edges will not have a significant impact on the growth rate of dynamical instabilities \citep{pittard2009turbulent}.

\begin{figure}
\vspace*{-0.5cm}
\includegraphics[width=0.5\textwidth]{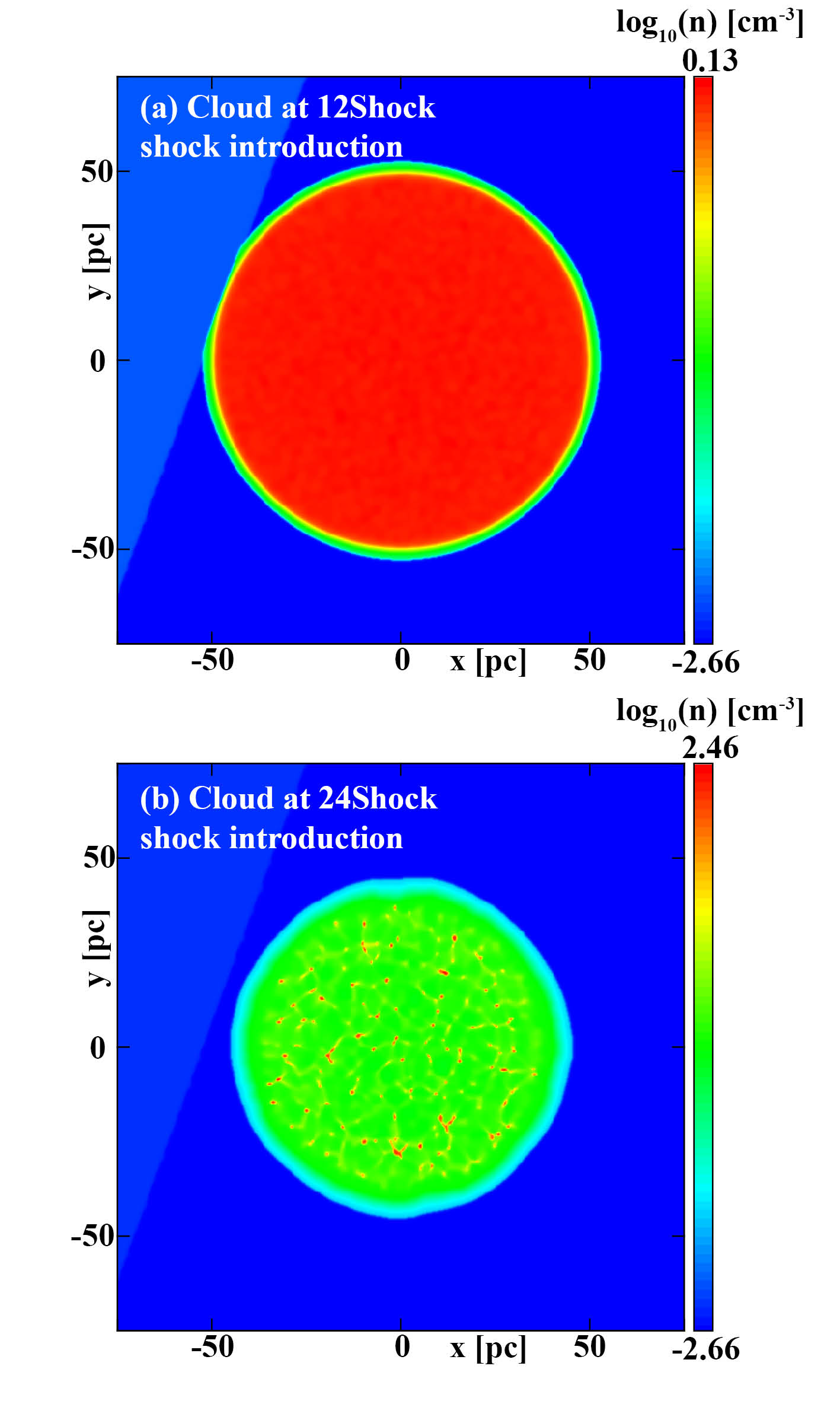}
\caption{(a): Cloud at shock introduction for \textit{12Shock} case after $\sim$\,12\,Myrs of evolution. (b): Cloud at shock introduction for \textit{24Shock} case after $\sim$\,24\,Myrs of evolution.}
\end{figure}

\subsubsection{Resolution and thermal instability}

The model employs 8 levels of AMR where the coarsest grid, \textit{G}$_0$, was set with a small number of cells (8$^3$) to ensure fast convergence of the multigrid solver. This meant that if fully populated, the finest grid \textit{G}$_7$ would employ 1024$^3$ cells resulting in an effective resolution of 0.29\,pc. Note that the shocked simulations also include a further single level of AMR for the \textit{12Shock} case and two extra levels of AMR for the \textit{24Shock} case towards their final stages, as they reach the resolution limit quickly.

\citet[][]{koyama2004field} assert that for simulations of the TI to converge, one must include thermal conduction and resolve the Field length by at least 3 cells, which they call the `Field condition': we do neither of those things. According to their equation (11) the Field length for our initial unperturbed density is 0.0587\,pc. We note however that their equation (11) is incorrect, and a more appropriate expression for the Field length was derived by \citet[][]{falle2020thermal} and is included in Appendix A. This expression gives a Field length of 0.594\,pc for our unperturbed density, so we would require closer to 20 cells. In any case, to satisfy this we would require excessive computational resources. However, \citet[][]{falle2020thermal} showed that with thermal conduction for our initial density \textit{n}\,=\,1.1\,cm$^{-3}$, rather than being narrowly peaked, the growth rate has a rather broad maximum between the Field length and the thermal length located at $\lambda$\,=\,8.95\,pc. Thus the TI does not depend strongly on increases in resolution within this range. In all of our simulations, perturbations grow initially at the grid scale but rapidly increase beyond that scale to large clumps. Although we exclude thermal conduction, these are nevertheless separated by length scales roughly corresponding to the wavelength with the maximum growth rate which is in common with various other authors who, despite insufficient resolution for the Field length, recover the properties of a thermally bistable medium -- it is accepted that such properties as the mass function of cold clumps and the power spectra of velocity and density are known to converge on large scale \citep{gazol2005pressure,vazquez2006molecular,hennebelle2007structure,inoue2015thermal}. Since our intention is to examine the large-scale interaction of a shock with a cloud, we argue that it is sufficiently resolved in this paper. We explore this further and perform resolution tests in Appendix A.

Our grid is set up so that 3 levels fully populate the domain and 8 levels fully populate the cloud region. This is in contrast to WPFVL16 who employed 5 fully populated levels with the remaining 3 refining and de-refining dynamically. Our choice is so that we are able to track all of the dynamics within the cloud when the shock is introduced and for numerical consistency, we keep this setting on throughout all of the cloud's evolution. It also allows us to extend the size of the domain from the -1.5 < \textit{xyz} < 1.5 used in WPFVL16 at little computational cost; placing the grid boundaries further from the cloud reduces the possibility of the cloud advecting off the grid and being affected by any shock induced reflected waves at the boundaries. It is important to note that fully refining the cloud does affect the initial behaviour of the TI, and so our evolution deviates slightly from WPFVL16. As the refinement of their highest 3 levels was not fixed, it resulted in their AMR grid de-refining to the 5th level as the density variations seeded in the initial condition smoothed out. As our inital perturbations smooth out, our grid does not de-refine and thus the growth rate at our highest level is larger than theirs and the TI develops earlier. This results in a difference in evolution time-scales, where their model experiences a delayed phase transition when compared to ours, and is seen to evolve for approximately a full analytical free-fall time of $\sim$\,50\,Myrs before reaching the state that we do at $\sim$\,35\,Myrs. We stress again however that while the time-scales are different, the final established state of the TI in terms of number, mass distribution and separation of clumps, is unaffected by this.


\subsection{Cases}


There are three cases studied in this work. The first case \textit{NoShock}, follows the evolution of the quiescent cloud described in the previous section. This is left to evolve for a free-fall time and used to compare against cases with shocks. In the second case \textit{12Shock}, the quiescent cloud is evolved for $\sim$\,12\,Myrs and then a shock is introduced. In the third case \textit{24Shock}, the quiescent cloud is left to evolve for $\sim$\,24\,Myrs before a shock is added. Note that the \textit{NoShock} case is evolved using the Godunov solver until the gradients become too large; it is then changed to Kurganov-Tadmor. This has a small effect on the mass distribution, however it eventually gets reconciled and it does not effect the global evolution of the cloud. For the shocked cases  we switch to Kurganov-Tadmor immediately prior to shock introduction, as the gradients introduced by the shock require a more diffusive solver.

The shock is artificially imposed on the grid by setting the values of cells to the left of the shock front according to the Rankine-Hugoniot relations. The shock then propagates from left to right. Note that cells that lie on the boundary in the post-shock region are also set to the post-shock values. The shock front is defined by the normal vector which lies at \textbf{\textit{r}}\,=\,1.05$\hat{\textit{r}}$ and $\theta$\,=\,160$^{\circ}$ where \textbf{\textit{r}} is the radial vector from the origin (where unit length\,=\,50\,pc) and $\theta$ is the angle counterclockwise from the positive \textit{x}-axis. This angle was chosen as simulations with the shock propagating directly along the \textit{x}-axis saw the formation of artificial structures. This was due to the Quirk instability \citep{quirk1992icase} which commonly arises in upwind schemes when the angle between the flow and grid axes is small. After the shock has been imposed it is evolved for a few short time-steps in order for the AMR grid to refine to the finest level at the shock front in regions where it is not fully refined. Following this, it is reimposed tangent to the cloud. In both cases, the shock has a Mach number of $\mathcal{M}$\,=\,1.5, where $\mathcal{M}$\,=\,\textit{v}$_{\textup{s}}$/\textit{c} for a shock with velocity \textit{v}$_{\textup{s}}$ moving into a stationary material with a sound speed \textit{c}. The associated clouds at those times are shown in Fig.\,2, which can be considered as the initial conditions for the shocked cases. The simulations are then evolved for a further 5.16\,Myrs and then stopped as densities reach values that are beyond the resolution limit set by the Truelove criterion \citep{truelove1997jeans}. The resolution is then increased by allowing additional grid levels and the simulation is evolved for an additional 1.5\,Myrs.

We follow the small-cloud approximation of \citet[][]{klein1994hydrodynamic}, which requires that a cloud with radius \textit{r}$_{\textup{cl}}$ and a shock-source blast wave of radius \textit{R}$_\textup{blast}$ satisfies \textit{r}$_{\textup{cl}}$ $\ll$ \textit{R}$_\textup{blast}$. This assumption means that the shock introduced can be approximated as planar, making it the simplest approximation of a more complex interaction, e.g. between a cloud and a supernova remnant in the Sedov-Taylor stage. 


\begin{figure*}
  \includegraphics[width=0.9\textwidth]{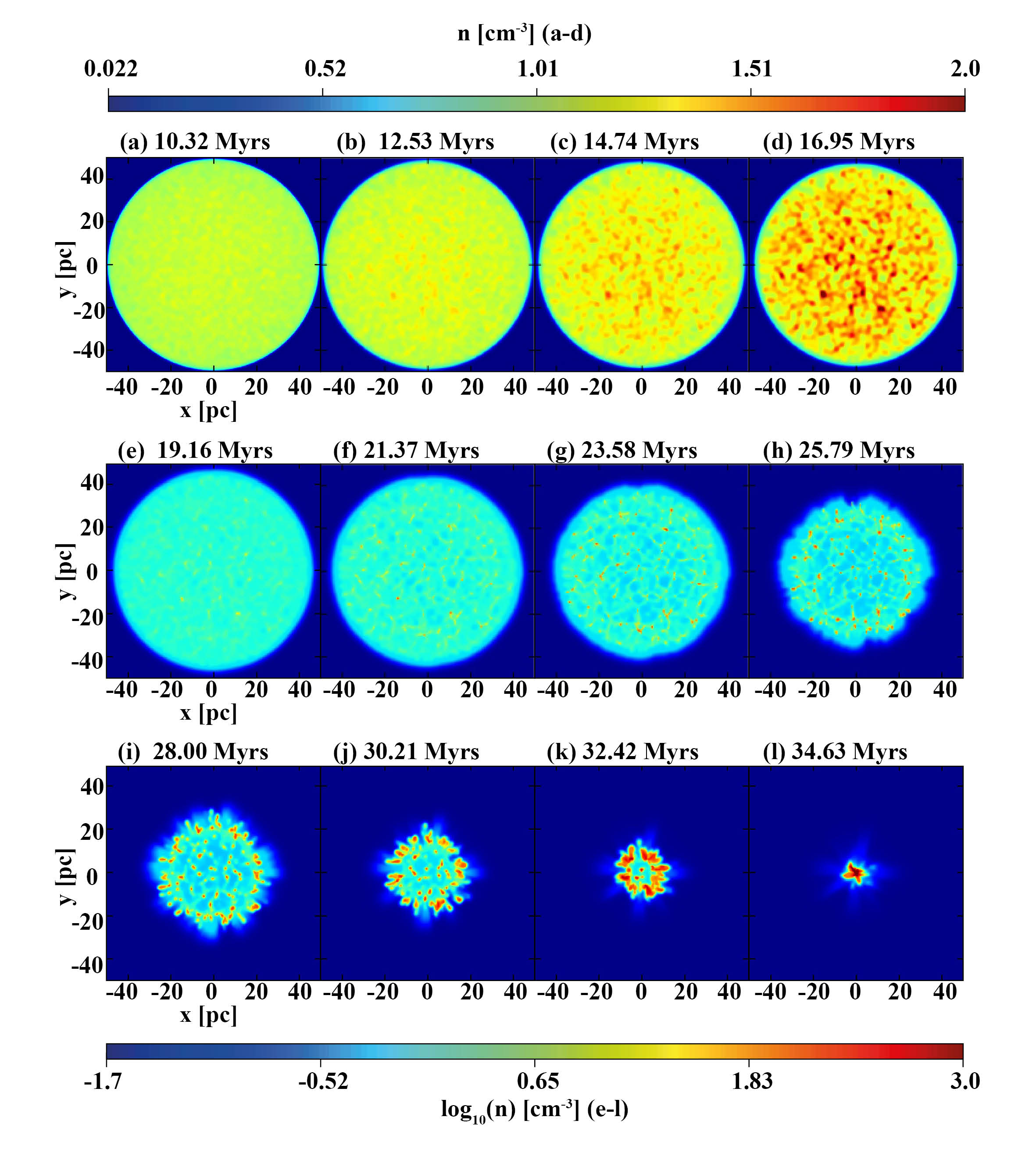}
  \caption{Slices through $\textit{z}$\,=\,0 of the \textit{NoShock} simulation evolving over a period of $\sim$\,15\,Myrs. Minimal change is seen in the cloud from \textit{t}\,=\,0 to 10.32\,Myrs and hence these snapshots are not shown. Snapshots highlight the onset of the instability (a\,--\,d), its development into a 2-phase medium (e\,--\,h), and the final collapse of the cloud (i\,--\,l).}
\end{figure*}

Even though the clouds at 12 and 24\,Myrs are very different from one another, it is useful to consider their evolution in terms of established theoretical time-scales and we follow the definitions of \citet{klein1994hydrodynamic}. For a cloud with radius $\textit{r}_{\textup{cl}}$ and an external shock with velocity $\textit{v}_{\textup{s}}$, the time taken for the external shock to sweep across the cloud, the \textit{shock crossing} time-scale, is defined as

\begin{equation}
    t_{\textup{sc}} \equiv \frac{2r_{\textup{cl}}}{v_\textup{s}}.
\end{equation}

\noindent During this period, a shock is driven into the cloud which propagates approximately with a velocity 

\begin{equation}
    v_{\textup{s,cl}} \equiv \frac{v_{\textup{s}}}{\chi^{1/2}}.
\end{equation}

\noindent The characteristic time for the cloud to be crushed by the transmitted shock, i.e the \textit{cloud crushing} time-scale, can now be defined as \textit{r}$_{\textup{cl}}$/\textit{v}$_{\textup{s,cl}}$, or as

\begin{equation}
    \textit{t}_\textup{cc} \equiv \frac{\chi^{1/2}\textit{r}_\textup{cl}}{\textit{v}_\textup{s}}.
\end{equation}

\noindent These are the basic time-scales governing the evolution of the shocked cloud. For our models with $\chi$\,=\,500, $\mathcal{M}$\,=\,1.5 and \textit{r}$_{\textup{cl}}$\,=\,50\,pc, we have \textit{v}$_{\textup{s}}$\,=\,240\,km\,s$^{-1}$, \textit{t}$_{\textup{sc}}$\,=\,0.41\,Myrs, \textit{v}$_{\textup{s,cl}}$\,=\,10.7\,km\,s$^{-1}$ and \textit{t}$_\textup{cc}$\,=\,4.6\,Myrs. 

It is important to note that these values are approximate and are derived for an adiabatic shock in a uniform cloud. While our shock is adiabatic in the ambient medium, inside the cloud it is radiative, and for model \textit{24Shock} there is structure inside the cloud. Nevertheless we find that the shock in our models still propagates on this time-scale.

\section{Results and discussion}
In this section we first present the \textit{NoShock} case. Following that, the evolution of the \textit{12Shock} and \textit{24Shock} cases are described. For the \textit{NoShock} case the times quoted are since \textit{t}\,=\,0. For the shocked cases the first times quoted are those elapsed since shock introduction and the corresponding times since \textit{t}\,=\,0 are quoted in brackets.

\subsection{Case 1 -- \textit{NoShock}}

\begin{figure*}
  \includegraphics[width=0.95\textwidth]{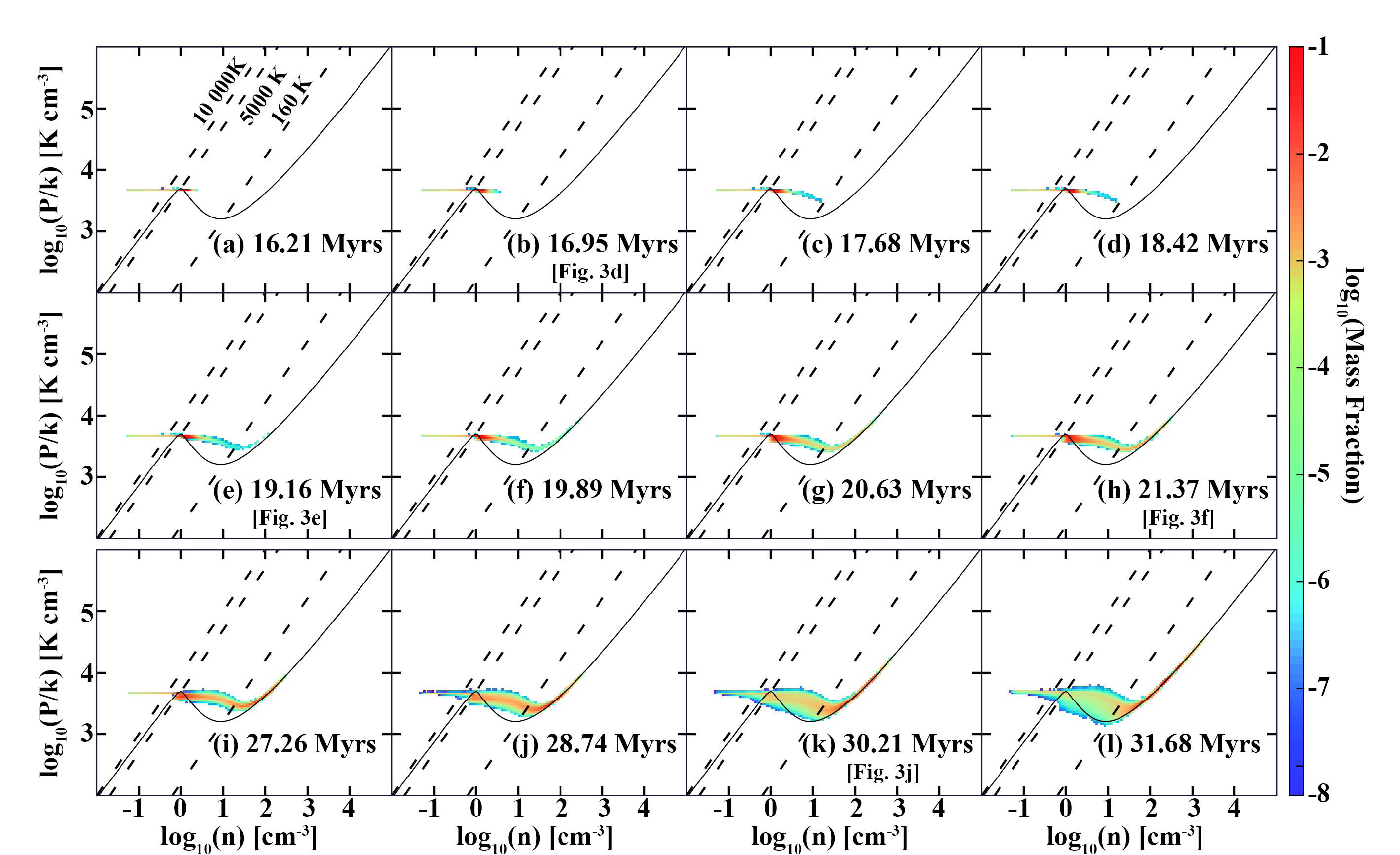}
 \caption{Mass distribution in pressure-density space from the moment of phase transition to a 2-phase medium. Panels (i\,--\,l) show the final stages of evolution as the cloud experiences global gravitational collapse. Temperatures that distinguish the different equilibrium phases are shown with isotherms that mark the hot phase (10\,000\,K\,$<$\,$\textit{T}$), warm phase (5000\,K\,$<$\,$\textit{T}$\,$<$\,10 000\,K), unstable phase (160\,K\,$<$\,$\textit{T}$\,$<$\,5000\,K) and cold phase ($\textit{T}$\,$<$\,160\,K). The corresponding panels in Fig.\,3 are referenced in square brackets where applicable.}
\end{figure*}

Fig.\,3 shows the evolution of the \textit{NoShock} case in density slices in the \textit{x\,--\,y} plane (slice through \textit{z}\,=\,0). Prior to Fig.\,3(a) at 10.32\,Myrs the cloud remains effectively unchanged with all gas initially found in the thermally unstable phase. The instability is seeded by the pressure variation across the cloud which grows initially on the smallest scales. The material in the cloud must then evolve into a thermally stable phase, where either it condenses into a cold, dense phase or evaporates into a warm diffuse phase. This results in the characteristic 2-phase medium of \citet[][]{field1969cosmic} which is seen to develop here. Fig.\,3(a\,--\,d) capture the first moments of the phase transition from the initially quiescent cloud. As there are no external influences on the cloud, the long period of quiescence up to this point reflects the growth rate of the TI at this resolution, which is relatively long but nevertheless shorter than the gravitational time-scale. 


Number densities of $\sim$\,100\,cm$^{-3}$ are first seen around 18\,Myrs, which are complemented with drops in temperature by nearly 2 orders of magnitude from $\sim$\,4000\,K to 100\,K. Fig.\,3(e) shows the first instance where a small number of higher density locations have emerged with clear, lower density structures spread throughout the cloud. These inhomogeneities are seen to develop into clumps at different rates, with some clumps clearly growing in mass and size faster than others. By 21.37\,Myrs (Fig.\,3f) there are a large number of clumps visible which are approximately separated at distances of 10\,--\,15\,pc. The average separation decreases to $\sim$\,5\,--\,10\,pc as more condensations form out of the material transitioning from the thermally unstable phase to the cold phase. This is the typical length scale of the TI-driven large-scale stable structures. By 23.58\,Myrs (Fig.\,3g) the cloud has settled to a state that is typical of a 2-phase medium formed due to the TI, which is seen to form in all hydrodynamic scenarios in WPFVL16. It is important to note that this spatial distribution and growth rate of clumps are characteristics of the TI, which are not affected by the spherical nature of the cloud or any edge effects. By mass, $\sim$\,60 per cent of the material is in cold and dense clumps ($\textit{T}$\,$<$\,160\,K) with $\sim$\,35  per cent of the material still unstable (160 $<$\,\textit{T}\,$<$ 5000\,K) located in a thin layer around the clumps. The majority of the volume is occupied by warm gas, however this accounts for only $\sim$\,5 per cent of the mass. 

Panels (i\,--\,l) in Fig.\,3 show the final evolutionary stages of the collapse. By 28\,Myrs the cloud has shrunk to a radius $\sim$\,20\,pc. At this point gravity is dominating over the TI in the evolution of the internal structures as the cloud is now accelerating rapidly at the edges, and as it collapses the outer clumps collide and merge with many of the inner clumps. This accelerates the growth of the maximum density which exceeds 1000\,cm$^{-3}$ by 30\,Myrs. By 32.42\,Myrs (Fig.\,3k) there is residue of the thermal instability formed clumps near the centre of the cloud, and the now 10\,pc radius cloud contains a cold and dense canopy all around the edges. This period is very short lived as by $\sim$\,35\,Myrs (Fig.\,3l) the whole cloud has collapsed to a single core with radius 5\,pc containing $\sim$\,17\,000 M$_{\odot}$ material.

Fig.\,4 shows snapshots of the evolution of the pressure-density mass distribution imposed over the pressure equilibrium curve. Isotherms are shown to distinguish between the hot, warm, unstable and cold phases. It is important to note that only material lying on the equilibrium curve can be considered to belong to a phase, otherwise it is simply transitioning within a regime or between phases. Any gas that does not lie on the equilibrium curve but is found in a temperature regime is explicitly referred to as that regime's material, e.g. unstable \textit{material}, which lies in the unstable temperature range but is not on the equilibrium curve.

Little change is seen in the gas state between \textit{t}\,=\,0 (Fig.\,1b) and 16\,Myrs (Fig.\,4a) where we can see the majority of the gas in the cloud is still located in the unstable phase. Note that we have a smoothed cloud interface, which is responsible for the distribution of material seen in Fig.\,4(a). On this time-scale however, even clouds that start off with sharp edges develop a smoothed edge due to the solver distributing discontinuities over a number of grid cells. Although little change since \textit{t}\,=\,0 is seen in Fig.\,4(a), it does capture the first hints of the phase transition due to TI. In Fig.\,4(a\,--\,c) material is seen to migrate across the phase diagram and by 18.42\,Myrs there is a fraction of cold material settling into the cold phase. This is reflective of the time-scale of the TI and the formation of the first clump. Following this we see the migration of more unstable material into the cold phase, which manifests as an increase in the number of clumps and mass in individual clumps. This behaviour continues throughout the cloud until \textit{t}\,$\approx$\,29\,Myrs. Fig.\,4(k) is when we start to observe the effects of the merging and coalescence increasing the maximum density, with gravity keeping the structures bound together. At \textit{t}\,$\approx$\,32\,Myrs (Fig.\,4l), 95 per cent of mass is now contained in the cold phase with remaining material being warm and unstable and the gravitational collapse of the whole cloud follows.

Two behaviours are important to note from the phase diagrams. The first is the migration of the unstable material into the cold phase as seen in Fig.\,4(a\,--\,c). This type of transition is the precise signature of the TI when viewed on a pressure-density phase diagram. If a shock is introduced into a simulation prior to this, and shocked gas cools into the unstable phase it is then susceptible to the TI \citep[e.g.][]{van2010shock}. If it consequently follows this trajectory, one can conclude that the shock has successfully triggered the TI \citep[e.g.][]{inoue2009two,inoue2016formation}. The second thing to note is that in isolation with no other influences, the final state of the TI from this initial condition can produce maximum cold densities of 100\,cm$^{-3}$. Gravity and cooling changes these slightly, however the persistence of the mass distribution as seen in Fig.\,4(h,i) represents what can be considered the final established state of the TI. We briefly highlight the spread of the distribution seen to occur between Fig.\,4(h/i). This is partly due to the changing of the Riemann solver from Godunov to Kurganov-Tadmor, which temporarily reduces the maximum density on the grid and diffuses the sharply peaked density profiles of the clumps. The 2-phase structure is not strongly affected by this, however large deviations from this distribution means the development of a different environment that may not closely resemble a 2-phase medium. The existence and deviations of such behaviour are some of the things examined in our shock-cloud interactions, which we now turn to.

\begin{figure*}
  \includegraphics[width=0.95\textwidth]{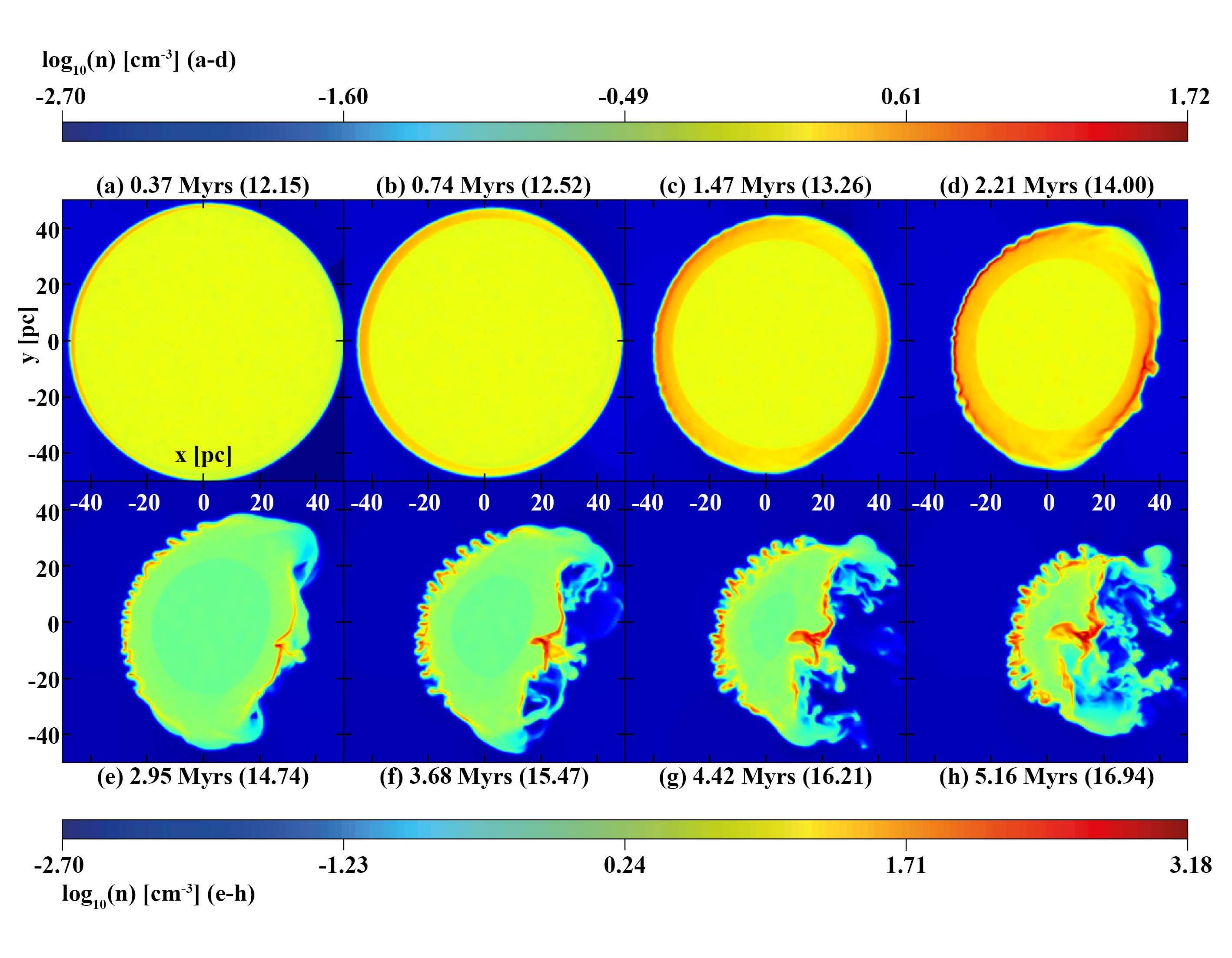}
  \caption{Slices through $\textit{z}$\,=\,0 of model \textit{12Shock} evolving over a period of 5.16\,Myrs taken every 0.74\,Myrs. The logarithm of the number density is shown with a separate colour scale for panels a\,--\,d and e\,--\,h. The time quoted first is the time elapsed since shock introduction and the time quoted in brackets is the time since \textit{t}\,=\,0. }
\end{figure*}


\subsection{Case 2 -- \textit{12Shock}}\label{sec:case2}
\subsubsection{Dynamics and morphology}

The evolution of the cloud in the \textit{12Shock} case is illustrated with density slices in the \textit{x\,--\,y} (\textit{z}\,=\,0) plane in Fig.\,5. 

Immediately prior to introducing the shock, the cloud has not experienced any significant changes. It has not yet contracted, its density, pressure and temperature are almost uniform and almost all of the gas is in the thermally unstable phase where it would have remained for another 4\,Myrs if undisturbed (as seen in Fig.\,4a). Initially, therefore, the behaviour has characteristics common to the shock-cloud interactions seen in many works and can be effectively described as having a constant density contrast of 500 and a theoretical cloud crushing time-scale of $\sim$\,4.6\,Myrs. Note that while the transmitted shock cools and therefore slows down, it is seen to converge in the cloud centre prior to the snapshot at 5.16\,Myrs (Fig.\,5h), which is of the order of the analytical cloud-crushing time-scale.

\begin{figure*}
  \includegraphics[width=0.95\textwidth]{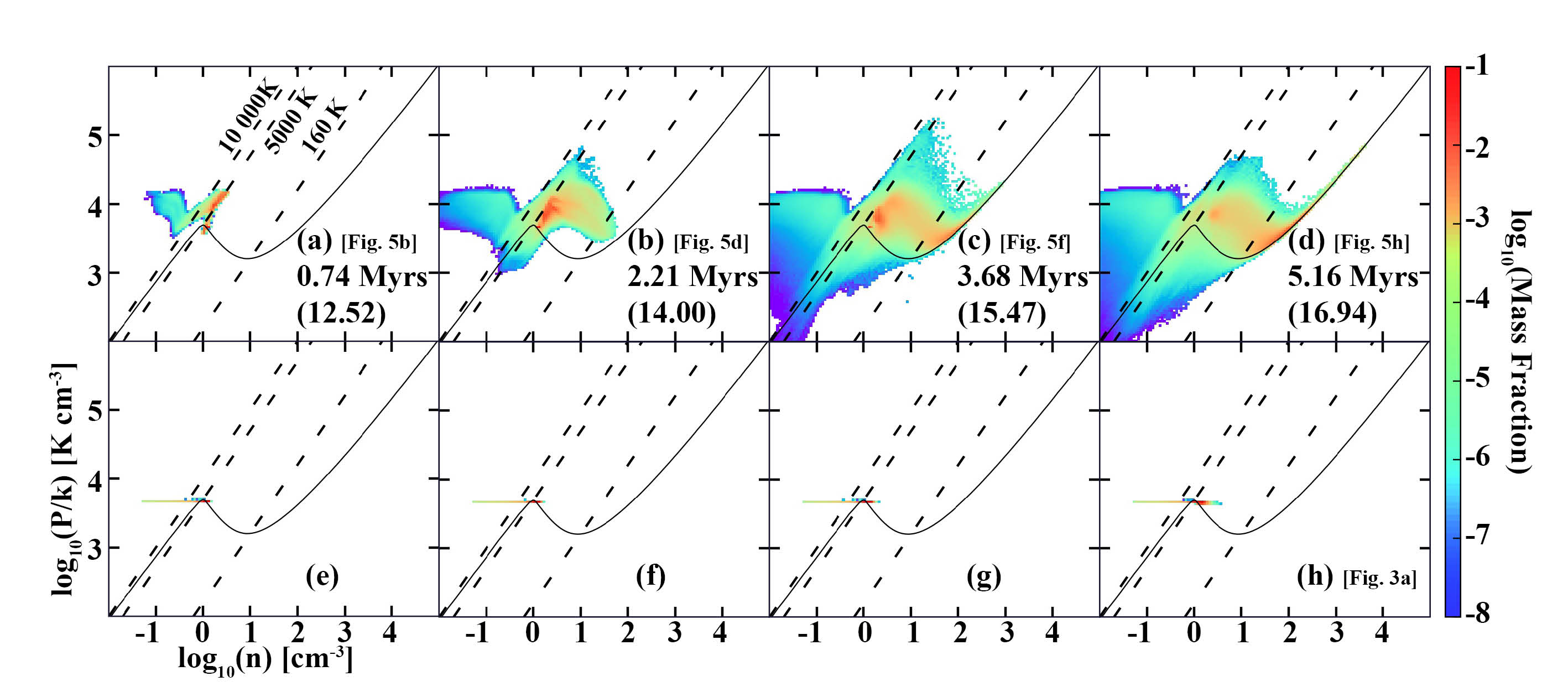}
  \caption{(a\,--\,d): Phase diagrams for model \textit{12Shock} over a period of 5.16\,Myrs taken every 1.48\,Myrs. (e\,--\,h): Phase diagrams for the corresponding snapshots in model \textit{NoShock}. Logarithmic mass-weighted mass fraction is shown. The first time quoted corresponds to elapsed time since shock introduction and the time in brackets is since \textit{t}\,=\,0. The corresponding density slices in Fig.\,5 (Fig.\,3 where applicable) are referenced in the square brackets. Isotherms delineating the hot, warm, unstable and cold regimes are shown.}
\end{figure*}

\begin{figure*}
  \includegraphics[width=0.95\textwidth]{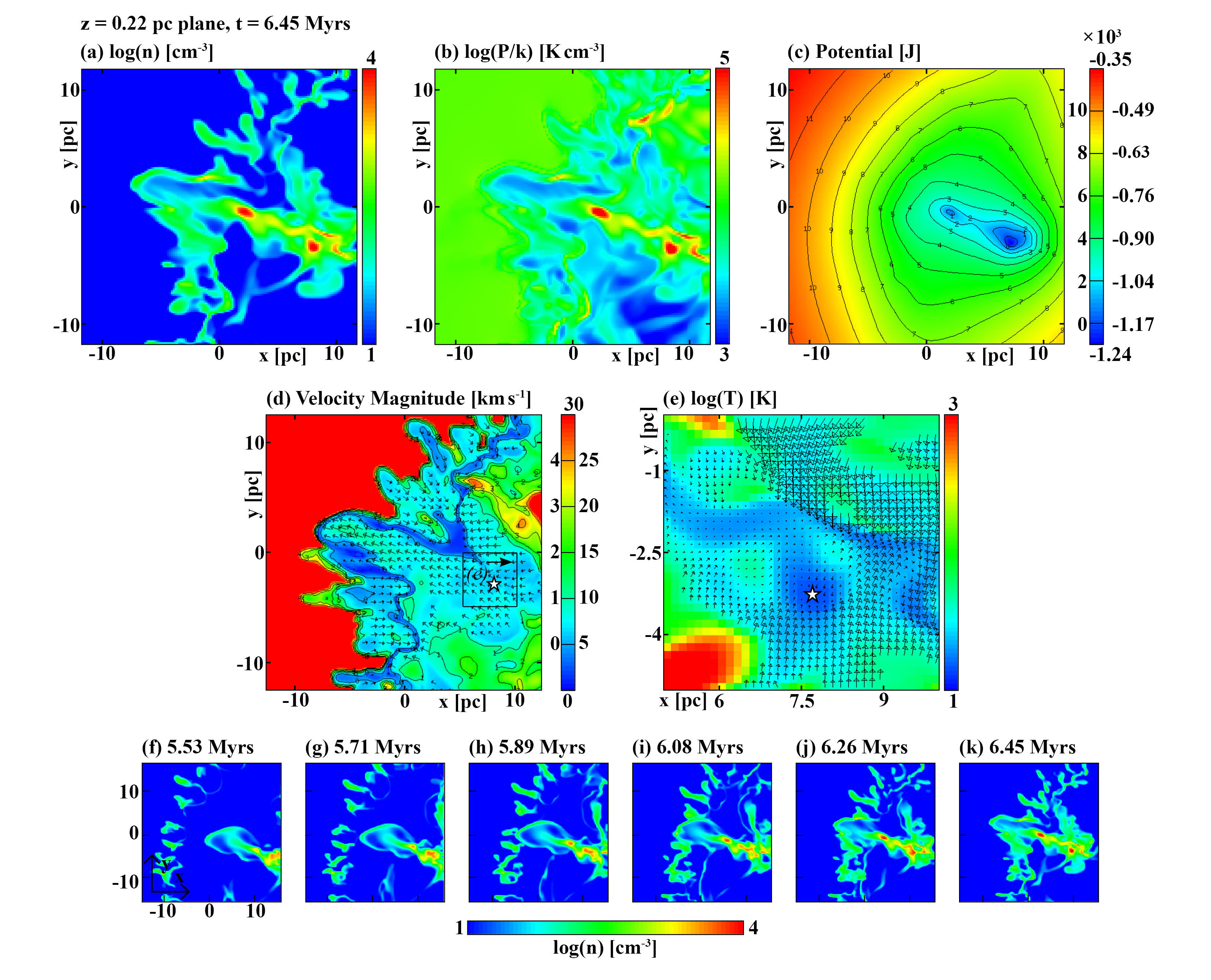}
  \caption{Collapsing clump in the \textit{12Shock} case shown at \textit{t}\,=\,6.45\,Myrs. The slice is through the region of maximum density ($\textit{z}$\,=\,0.22\,pc). Shown are (a) density, (b) pressure and (c) gravitational potential. Panel (d) shows magnitude of velocity. Velocity vectors are shown only for gas in the cold ($\textit{T}$\,$<$\,160\,K) regime scaled to \textit{v}\,=\,10.84\,km\,s$^{-1}$, the largest velocity in the region constrained by the temperature. Panel (e) shows a region zoomed in on the collapsing core with velocity vectors in the frame of the clump for the coldest ($\textit{T}$\,$<$\,50\,K) regions showing a converging velocity field. Velocity vectors are scaled by the largest velocity in this region which is \textit{v}\,=\,6.12\,km\,s$^{-1}$. Panels (f\,--\,k) show the density evolution up to that point.}
\end{figure*}

The external shock has a velocity of 240\,km\,s$^{-1}$ and crosses the cloud in less than 0.45\,Myrs. During the passage of the external shock, a bow wave is reflected back into the external medium and a shock is transmitted into the cloud. In this scenario, \textit{t}$_{\textup{cc}}$/\textit{t}$_{\textup{sc}}$ $\sim$\,10 so the transmitted shock propagates much slower than the external shock. The external shock is diffracted as it sweeps over the cloud causing it to lose strength. Swept up material is therefore raised to a lower pressure. This results in a weaker transmitted shock further down the cloud which can be seen in the decrease of the initial density jump on the sides in Fig.\,5(b) when compared to the initial density jump on the front face of the cloud in Fig.\,5(a). Consequently the cloud size is mainly reduced in the direction parallel to the distant upstream flow velocity. The ram pressure of the external flow accelerates this process. When the external shock converges on the symmetry axis behind the cloud, a strong pressure increase occurs which drives a shock into the back of the cloud. The consequences of this are seen after 1.5\,Myrs in Fig.\,5(c) where the transmitted shock is clearly visible within the entire cloud with higher densities at the back and front when compared to the sides. This results in a cloud that is increasingly oblate.

At the front of the cloud, the transmitted post-shock gas cools on the order of a cooling time-scale (\textit{t}$_{\textup{cool}}$\,$\approx$\,1.5 Myr) forming a dense shell near the cloud interface. As the shell's internal energy is radiated away, its pressure drops and so it is further compressed making it susceptible to various dynamical instabilities. For example, the acceleration of the cloud by the post-shock gas triggers the Rayleigh-Taylor (RT) and Kelvin-Helmholtz (KH) instabilities. The RT instability in general arises as a consequences of pressure and density differences between colliding flows or within static stratified media, and hence here is most disruptive on the front face of the cloud. At 2.21\,Myrs (Fig.\,5d), regions in the shell are seen to accelerate at different rates resulting in larger, more distinct perturbations. These grow to resemble RT `fingers' which distort the shell and make it susceptible to more unstable behaviour. Between 3\,--\,4\,Myrs (Fig.\,5e\,--\,f) these distortions have grown so large that the instability now resembles the Vishniac instability \citep{vishniac1983dynamic,michaut2012numerical}, strikingly similar to what was witnessed by \citet[][]{pittard2003formation} in an expanding SNR in an AGN environment. 

The KH instability is most disruptive in regions with large velocity shear, which in this work is on the edges of the cloud. Where there is shear, small-scale perturbations are amplified to lateral and vertical motions that form locally circulating eddies that disrupt the cloud. At the rear of the cloud, the velocity shear results in the formation of a powerful poloidal flow with a low pressure core, which redirects the external stream adjacent to the cloud towards the back of the cloud. This results in a vortex ring with a complex velocity field that traps hot circulating low-density gas and induces strong turbulent motions causing mixing of cloud and ambient material. This flow warps the outer edges of the rear-facing shell and amplifies the compression of the already convergent flow. This strongly affects the morphology of the shell which in the density slices is seen as a winged-like structure with a protruding needle at the centre, where the flow is most focused. The growth of this structure is seen most clearly in Fig.\,5(e\,--\,h) where the strong pressure jump and converging flow induces velocities of up to 13\,km\,s$^{-1}$ in the needle, forcing it to grow in contrary motion to the 1.8\,km\,s$^{-1}$ motion of the centre of mass of the cloud. This is also faster than the transmitted shock velocity of 10.6\,km\,s$^{-1}$ and while not apparent in the colour scale used here, the needle overshoots the transmitted shock by Fig.\,5(g). By Fig.\,5(h) the cloud appears hemispheric with a curved front and a flattened back. The exterior contains multiple clumps and dense structures with a needle like protrusion nested at the rear. The base of the needle contains a high density core with \textit{n}\,=\,1.51\,$\times$10$^{3}$\,cm$^{-3}$.

\subsubsection{Phase evolution}

The evolution of the mass weighted pressure-density distributions are shown in Fig.\,6. Panels (a\,--\,d) show the distributions for the \textit{12Shock} case which are contrasted against the distributions of the \textit{NoShock} case in panels (e\,--\,h) for the same instances in time. The rich complexity of the interaction is indeed reflected in the plots. Initially, gas is seen to be shocked off the equilibrium curve increasing the amount of warm gas in the cloud. The full range of compression is captured, reflecting the variation of the strength of the transmitted shock across the cloud. The peak of the compression corresponds to regions in the front of the cloud where the transmitted shock is the strongest and gas cools the fastest. This gas loses much of its internal energy and thus is susceptible to further compression. This causes a significant fraction of unstable material to transition to the cold regime.

One focus of this work is to examine the interplay between the shock-dominated effects and the TI. Two things are important to highlight. When the shock is first transmitted it compresses the material such that it begins to cool rapidly. This results in rapid compression and material is seen to completely pass through the unstable regime. During this transition, it is not actually evolving due to the TI but radiative cooling. As mentioned in Section 3.1, the trajectory of the TI across the equilibrium curve is very distinct and evolves almost isobarically, which is not the case in this simulation. There does appear to be a possibility of re-triggering the TI, with some shocked gas transitioning, or remaining close to the unstable equilibrium curve. If this material was to settle in the unstable phase, it would again become subject to the TI. However following the convergence of the transmitted shock at the centre of the cloud on the cloud crushing time-scale (Fig.\,5f\,--\,h), all gas in the unstable phase gets shocked out of equilibrium. This gas is found where the transmitted shock was the weakest on the edges of the cloud. Therefore a weaker shock and a delayed cloud crushing would aid the development of the TI. A weaker shock could be achieved if the initial Mach number is lower, or if magnetic fields are present \cite[e.g.][]{van2010shock}. A larger cloud like that of WFP19 would also delay the cloud-crushing. The most effective shocks in triggering the thermal instability are those where the density of the final state is on the unstable part of the equilibrium curve. For an initial upstream density of \textit{n}\,=\,0.1\,cm$^{-3}$, \citet{falle2020thermal} show the path of an oblique MHD fast shock in their Fig.\,7 which achieves precisely this. See also discussion in their section 3.3 and further examples of such shocks in their table 1.

Densities greater than 1000\,cm$^{-3}$ are seen 4.42\,Myrs into the evolution of \textit{12Shock}. Contrasting this with the values seen in \textit{NoShock}, densities greater than 1000\,cm$^{-3}$ are first seen $\sim$\,28\,Myrs into its evolution and $\sim$\,15\,Myrs later than in the \textit{12Shock} case. Note that since the state of the cloud in model \textit{NoShock} is effectively the same over the first 15\,Myrs of evolution, introducing a shock at any time during this period will result in the same interaction as \textit{12Shock}. This means that densities of $\sim$\,1000\,cm$^{-3}$ could have been witnessed as much as 25\,Myrs prior to what is seen in the \textit{NoShock} case had we decided to introduce the shock at \textit{t}\,=\,0, for example. This long period of quiescence means that the likelihood of this type of interaction taking place in the ISM is therefore relatively high. In this scenario, the TI does not play as much of a role as the shock and gravity do. However, the cooling processes responsible for the TI are fundamental in this interaction.

\subsubsection{Local and global collapse of 12Shock}

We now discuss the state of the clumps that have formed in the context of star formation, compare this to what we see in the unshocked scenario \textit{NoShock}, and present evidence for local gravitational collapse. The evolution presented in the previous section ran to its resolution limit and to study any further evolution would require additional levels of AMR. WFP19 performed higher resolution simulations of a smaller portion of a larger cloud, one of which was focused on the gravitational collapse of an individual clump. This clump witnessed a rise of density and pressure by two orders of magnitude in $\sim$\,10$^5$ years. To track the entirety of this behaviour in our shocked clouds would require at least an additional 4 levels of AMR to increase the resolution from $\Delta$\textit{x}\,=\,0.29\,pc to $\Delta$\textit{x}\,=\,0.018\,pc as was done to capture the final collapse in WFP19. This would be extremely computationally intensive, so we choose to only add a single extra level of AMR and evolve slightly further. To discuss the differences between the structures that form, we follow the definitions summarised in table 1 of \citet{bergin2007cold} unless stated otherwise.

Prior to a cloud-crushing time-scale of $\sim$\,4.6\,Myrs, the cloud has formed a hemispheric, turbulent-like exterior which at the front face of the cloud fragments into structures that experience a density increase from \textit{n}\,=\,1.1\,cm$^{-3}$ to $\sim$\,200\,--\,1000\,cm$^{-3}$. In terms of mean density, size and velocity extent, these have properties typical of Bergin $\&$ Tafalla clouds. They do however fall on the lower end of the range of cloud sizes, with most structures having radii of $\sim$\,1\,--\,4\,pc, which falls closer to Bergin $\&$ Tafalla clumps. Their temperatures, crossing times and masses also agree with this definition. The larger clumps contain various substructures and appear to be coalescing with their neighbours. However their densities fail to grow much further beyond 1000\,cm$^{-3}$ and the clumps are not seen to collapse under gravity.  

At 6.63\,Myrs we find densities greater than 10$^4$\,cm$^{-3}$ and turn our attention to the structures containing them. The origin of these structures is witnessed earlier of course, as they are located at the base of the protrusion shown in Fig.\,5(f). In terms of sizes, masses and velocities, these structures continue to have properties that can be characterised as clumps. They now also fit the clump category according to their mean density, although the highest densities suggest that they are evolving into cores. In Fig.\,7 we show a snapshot of two of the clumps at this instance in time, 1.3\,Myrs of evolution after Fig.\,5(h). Evolution of this region up to this point is shown in Fig.\,7 panels (f\,--\,k). The slices presented are through the plane \textit{z}\,=\,0.22\,pc which cut through the location of maximum density which is \textit{n}\,=\,1.34\,$\times$10$^4$\,cm$^{-3}$ in the final snapshot; it is marked with a star on the velocity plot in Fig.\,7(d). We zoom in on this region in the temperature plot in Fig.\,7(e) and show locally converging velocity vectors in the frame of the collapsing clump. Note that velocity vectors in Fig.\,7(d) are shown for cold material ($\textit{T}$\,$<$\,160\,K), whilst in Fig.\,7(e) we only show them for the coldest ($\textit{T}$\,$<$\,50\,K) regions.

From the evolution snapshots in Fig.\,7(e\,--\,j), we see that the clumps are contained within the protrusion-like structure which is moving upstream at $\sim$\,8\,km\,s$^{-1}$ against the 90\,km\,s$^{-1}$ post-shock flow. One clump follows the motion of the needle and is seen to develop a local potential minimum. The second clump remains confined at the base of the potential well, but is accelerated upstream by the converging flow along with the material at the back and sides of the cloud. Its centre of mass has a velocity of 6.4\,km\,s$^{-1}$ and its internal velocity dispersion is $\sim$\,1.4\,km\,s$^{-1}$. At the front of the protrusion, a bow shock has formed which punctures through the front shell in Fig.\,7(k) and disrupts structures contained in the shell. While we notice regions get compressed and become over-pressured, on these time-scales and at this resolution, this interaction does not cause any structures in the shell to collapse.
\begin{figure*}
  \includegraphics[width=0.95\textwidth]{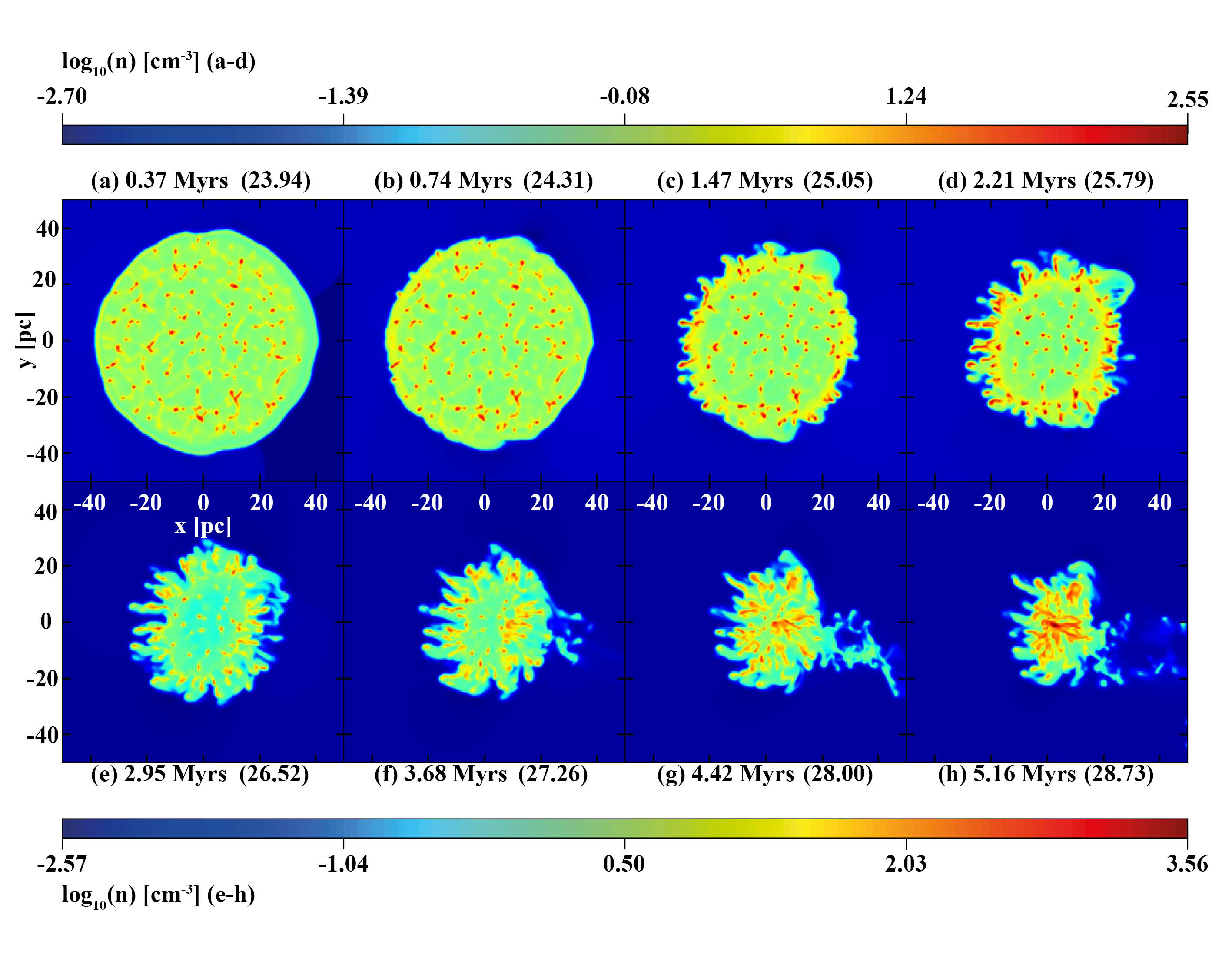}
  \caption{Slices through $\textit{z}$\,=\,0 of model \textit{24Shock} evolving over a period of 5.16\,Myrs taken every 0.74\,Myrs. The logarithm of the number density is shown with a separate colour scale for panels a\,--\,d and e\,--\,h. The time quoted first is the time elapsed since shock introduction and the time quoted in brackets is the time since \textit{t}\,=\,0.}
\end{figure*}

We focus therefore on the fastest collapsing clump marked by the star. \citet{williams1995density} found that in the Rosette Molecular Cloud, CO traced clumps have an average H$_2$ number density of \textit{n}\,$\approx$\,220\,cm$^{-3}$, excitation temperatures \textit{T}\,$<$\,20\,K and thermal gas pressures \textit{P/k}\,$\approx$\,2500\,K$\,$cm$^{-3}$. For our analysis we therefore isolate a spherical region centered on the maximum density with a radius set by inspection, and only trace material for \textit{n}\,$>$\,220\,cm$^{-3}$, \textit{P/k}\,$>$\,2500\,K\,cm$^{-3}$ and \textit{T}\,$<$\,20\,K. For the clump under consideration, our radius of choice is 0.75\,pc, which traces slightly over the total volume of the clump, and the other constraints ensure that effectively only molecular material is considered. We note that due to the low resolution, the location of the potential minimum, maximum density and the centre of mass of the core are all displaced by only a few cells and so do not affect out calculations.

We find that the clump contains $\sim$\,140 M$_{\odot}$ of material and has a mean density of \textit{n}\,$\approx$\,2.8\,$\times$10$^3$\,cm$^{-3}$. It has central temperatures of $\sim$\,12\,K and is over-pressurised with respect to the surroundings by almost 2 orders of magnitude. In the frame of reference of the clump, the surrounding velocities are converging and an energy analysis reveals it to be Jeans unstable and gravitationally bound. The density threshold for star particle creation as set by \citet{truelove1997jeans} is 1.1\,$\times$10$^4$\,cm$^{-3}$. As this has been exceeded and the usual tests for star particle algorithms have been passed \citep[e.g.][]{federrath2010modeling}, the clump could now have been converted to a star.

Since densities typical of molecular clouds were seen $\sim$\,2\,Myrs into the simulation, we conclude that the lifetime of our starless cloud, and therefore time-scale for star formation, determined by the first signs of gravitational collapse is $\sim$\,3\,--\,5\,Myrs. The application of a robust star particle algorithm to fully determine the star formation rates and efficiencies, and to study the feedback of those stars into these clouds, will be left for future work.


\subsection{Case 3 -- \textit{24Shock}}\label{sec:case3}


\begin{figure*}
  \includegraphics[width=0.95\textwidth]{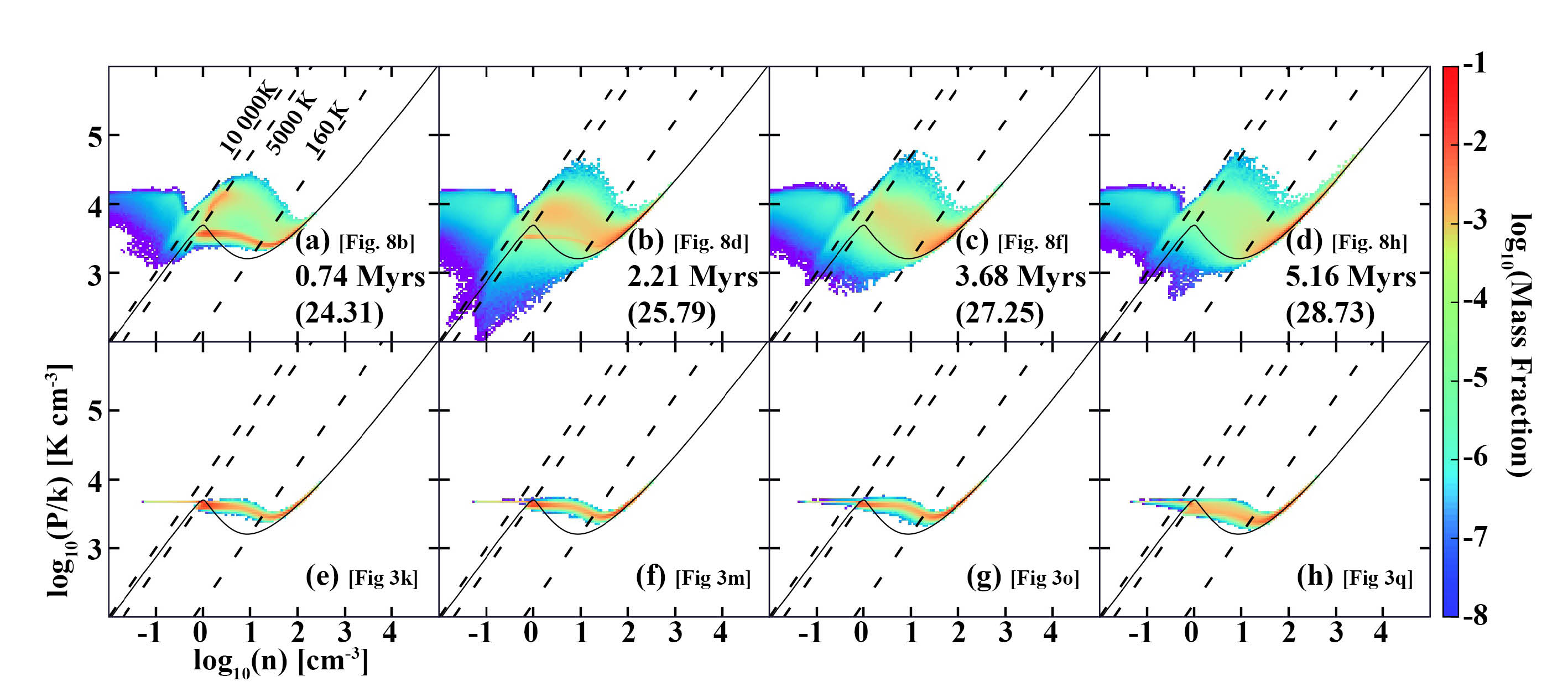}
  \caption{(a\,--\,d): Phase diagrams for model \textit{24Shock} over a period of 5.16\,Myrs taken every 1.48\,Myrs. (e\,--\,h): Phase diagrams for the corresponding snapshots in model \textit{NoShock}. Logarithmic mass-weighted mass fraction is shown. The first time quoted corresponds to elapsed time since shock introduction and the time in brackets is since \textit{t}\,=\,0. The corresponding density slices in Fig.\,8 (/3) are referenced in the square brackets. Isotherms delineating the hot ($\textit{T}$\,$>$\,10\,000\,K), warm (5000\,K\,$<$\,$\textit{T}$\,$<$\,10\,000\,K), unstable (160\,K\,$<$\,$\textit{T}$\,$<$\,5000\,K) and cold regimes ($\textit{T}$\,$<$\,160\,K) are shown.}
\end{figure*}

\begin{figure*}
  \includegraphics[width=0.95\textwidth]{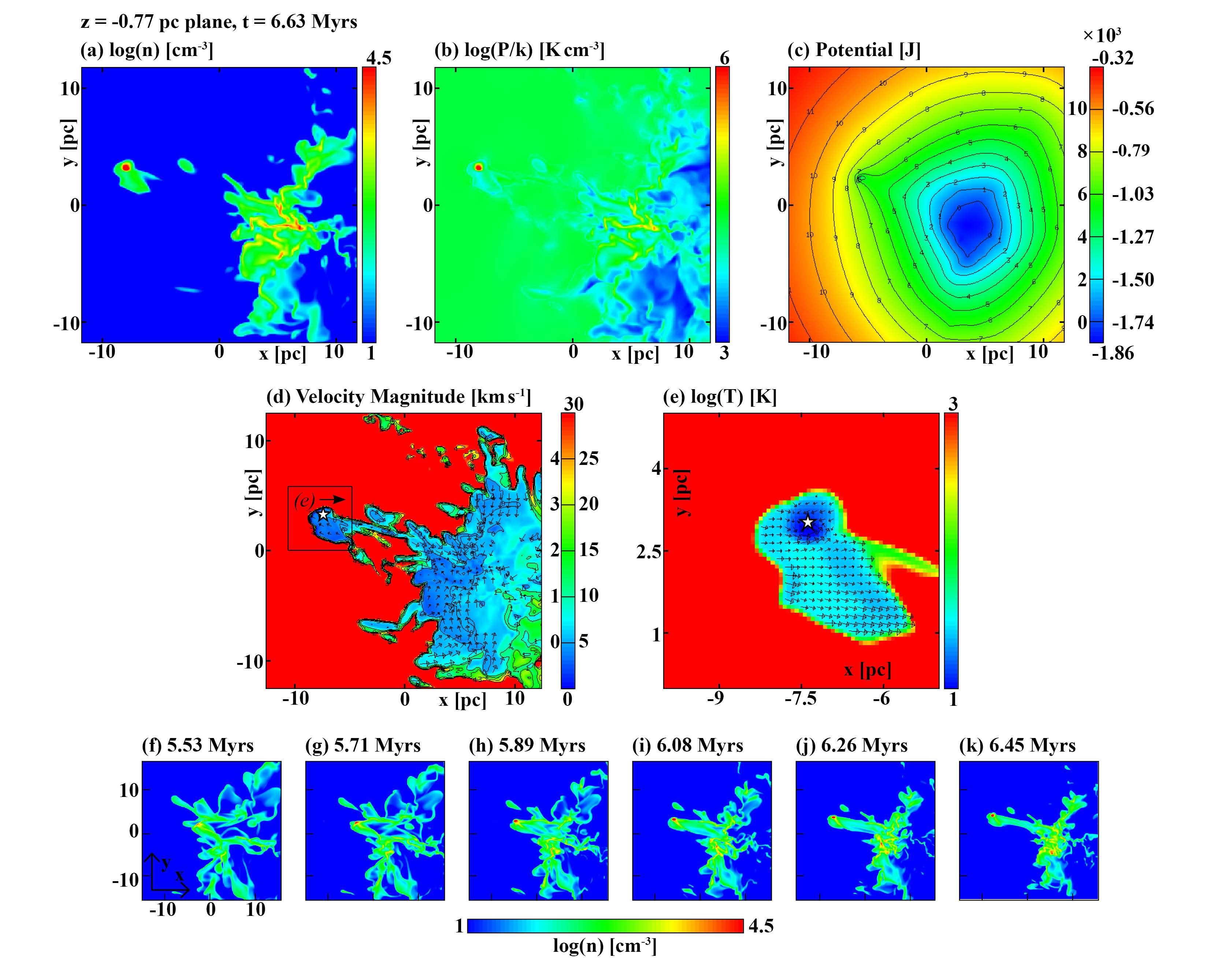}
  \caption{Collapsing clump in the \textit{24Shock} case shown at \textit{t}\,=\,6.63\,Myrs. The slice is through the region of maximum density ($\textit{z}$\,=\,-0.77\,pc). Shown are (a) the density, (b) pressure, (c) gravitational potential. Panel (d) shows the magnitude of velocity with velocity vectors for gas in the cold (\textit{T}\,$<$\,160\,K) regime scaled by \textit{v}\,=\,10.31\,km\,s$^{-1}$. Panel (e) zooms in on the collapsing core and shows velocity vectors in the frame of the clump for the coldest (\textit{T}\,$<$\,50\,K) regions. The vectors are scaled by \textit{v}\,=\,8.52\,km\,s$^{-1}$ and show a converging velocity field. Panels (f\,--\,k) show the density evolution up to that point.}
\end{figure*}

\subsubsection{Dynamics and morphology}

The evolution of the cloud in the \textit{24Shock} case is illustrated with density slices in the \textit{x\,--\,y }(\textit{z}\,=\,0) plane in Fig.\,8(a\,--\,h). 
In this scenario the cloud has been allowed to evolve undisturbed for 23.58\,Myrs. Over this period it has formed a 2-phase medium due to the effects of the TI, namely a complex of cold and dense clumps embedded in a warm diffuse inter-clump gas, which is itself embedded in a hot rarefied medium. The nature of the interaction is very different to \textit{12Shock}. Now each dense clump acts like its own individual cloud, effectively resulting in multiple shock-cloud interactions. The global interaction is dominated by the sum of these. This means that the global evolution deviates strongly from a typical shock-cloud interaction as the transmitted shock is distorted by the clumps and the development of the Vishniac, RT and KH instabilities are suppressed. These instabilities would likely be present on the individual clumps but they are not adequately resolved in these simulations. 

In Fig.\,8(a) the shock has swept over most the cloud and a transmitted shock is faintly seen. As there are no clumps exactly on the edges of the cloud (i.e all clumps are surrounded by an envelope of warm gas) the shock that they initially interact with is not the external shock but the transmitted inter-clump shock. The interactions are different from clump to clump, as the transmitted shock enters regions of varying density and density contrasts between the clump and interclump gas. The subsequent structure of the shock is then affected, as it loses energy as it propagates through the cloud and is distorted by the previous clumps it interacts with. 

The cloud experiences considerable disruption much earlier than the one in model \textit{12Shock}. As the cloud is more porous, the shock moves more rapidly through the low-density inter-clump material and accelerates the clumps as it does. The denser clumps are more resistant and are accelerated more slowly. As a result, they get left behind and become entrained in the low-density external flow. The shocked inter-clump material cools and forms a dense shell, with dense hubs occupied by the pre-existing clumps. Some clumps are seen to break off as early as $\sim$\,1.5\,Myrs (Fig.\,8c) and become entrained in the external flow. Inner clumps then become entrained in the shocked inter-clump flow, and all embedded clumps form elongated tails that are directed radially inward. The tails of the clumps inside the cloud become exaggerated when the external flow channels the lower density inter-clump material further into the cloud, exposing more inner clumps to the external flow.

As clumps are accelerated less than the inter-clump gas, channels form from all sides of the cloud which play an increasingly important role in redirecting the external flow. Since the inter-clump gas is less resistant to the flow, and the initial cloud is smaller than \textit{12Shock} at shock introduction, the shock can propagate faster through the cloud than in the \textit{12Shock} case. As a result, the cloud-crushing time-scale is shorter, and the transmitted shock converges at the centre of the cloud $\sim$\,3.7\,Myrs into the interaction (Fig.\,8f). Overall this results in turbulent-like dynamics. However, there is still large-scale order to the flow. 

Flow meeting the cloud at the sides now gets redirected upstream and meets the flow coming in from the rear. This flow is better able to accelerate the high density clumps with some accelerating directly along a collision path. Such collisions result in the formation of an over-dense region in Fig.\,8(g) which becomes the location of the highest density in the cloud. This region continues to gain bulk momentum and in Fig.\,8(h) it is seen to be confined to a clump at the centre of the cloud. Densities in excess of $\sim$\,3000\,cm$^{-3}$ are now located in this clump, suggesting that gravity is taking over its evolution.

\subsubsection{Phase evolution}

Fig.\,9 shows the evolution of the mass weighted pressure-density distributions, where panels (a\,--\,d) show the distributions for the \textit{24Shock} case and (e\,--\,h) for the \textit{NoShock} case for the same instances in time. 
From Fig.\,9(a) the first thing to note is that material with densities \textit{n}\,$\approx$\,1.1\,cm$^{-3}$ respond the strongest to the introduction of the shock. This corresponds to the shocking of the interclump material, and the distribution reflects the full range of compression it initially experiences. When out of equilibrium, the material subsequently behaves very similarly to the \textit{12Shock} case as it migrates from the unstable regime to the cold phase. This reflects the formation of the interclump shell that connects the outermost clumps. Compared to \textit{12Shock}, shocked material is much more widely distributed within the regimes, with most of it out of equilibrium and cooling to the cold phase. This distribution of material is aided by the shocking of the thermally unstable layers on the edges of clumps which significantly disrupts the distribution of the characteristic 2-phase medium seen to prevail in the \textit{NoShock} scenario. 

By 3.68\,Myrs (Fig.\,9c), this distribution has disappeared completely due to the transmitted shock converging at the centre of the cloud, and almost all material is out of equilibrium and cooling into the cold phase. By Fig.\,9(d), most of the material is concentrated in the cold phase and the maximum density is approaching \textit{n}\,=\,10$^4$\,cm$^{-3}$ reflecting the gravity dominated evolution resulting from the collision of structures near the rear of the cloud.

\subsubsection{Local and global collapse of 24Shock}

The maximum density in the cloud at the time of shock introduction is \textit{n}\,=\,332\,cm$^{-3}$. The free-fall collapse time-scale for this value is $\sim$\,2.6\,Myrs. Local structures do eventually collapse, though on a longer time-scale of $\sim$\,5\,Myrs. Note that similar behaviour was seen in WPFVL16 and WPF19, where an extended period of $\sim$\,15\,Myrs was seen prior to any collapse, in spite of a free-fall time of $\sim$\,5\,Myrs for \textit{n}\,=\,100\,cm$^{-3}$.

From the last snapshot shown in Fig.\,8(h), we continue the simulations with an additional 2 levels of AMR for a further 1.5\,Myrs. Fig.\,10 shows the final snapshot at 6.63\,Myrs where the slices are through the \textit{z}\,=\,-0.77\,pc plane and cuts through the fastest collapsing object marked by a star on the velocity and temperature plot in panels (d) and (e) respectively. Velocity vectors and the density evolution are shown as in Section 3.2.3. There continues to be a strong asymmetry between the momentum transferred from the front of the cloud when compared to the back and sides. Flow meeting the cloud at the sides is re-directed towards the centre by pre-existing structures where it picks up an upstream velocity component due to the interaction with the converging flow originating from the rear. 

We again analyse the fastest collapsing clump, which is located slightly off the central plane and in the head of the most protruding structure in the cloud. We analyse material using the same criteria as in Section 3.2.3 and find that the object contains 160 M$_{\odot}$ of material with a mean density of 1.47\,$\times$10$^4$\,cm$^{-3}$. It has bulk velocities $\sim$\,4\,km\,s$^{-3}$ in the ambient frame and an internal velocity dispersion of $\sim$\,1.38\,km\,s$^{-1}$. Fig.\,10(b) shows that it contains pressures in excess of 10$^5$\,K\,cm$^{-3}$ which are $\sim$\,1.5 orders of magnitude greater than the surroundings. With a developed local potential minimum and temperatures below 15\,K, this shows clear evidence of gravitational collapse. An energy analysis additionally reveals it to be Jeans unstable and gravitationally bound. The density threshold for star particle creation in the cell with highest density is 3.48\,$\times$10$^4$\,cm$^{-3}$ for this resolution. The maximum density in this object is \textit{n}\,=\,7.9\,$\times$10$^5$\,cm$^{-3}$, and so this object satisfies star particle formation criteria.

The lifetime for this cloud prior to star formation is slightly longer than in \textit{12Shock}, as we are also considering the period that established the 2-phase medium due to the TI. Densities in excess of 220\,cm$^{-3}$ were seen $\sim$\,18\,Myrs into the \textit{NoShock} model. Since the shock was introduced at $\sim$\,24\,Myrs and collapse happened 6\,Myrs after that, the lifetime of the starless molecular cloud, and therefore the time-scale for star-formation, is $\sim$\,12\,Myrs.

\begin{figure*}
  \includegraphics[width=0.95\textwidth]{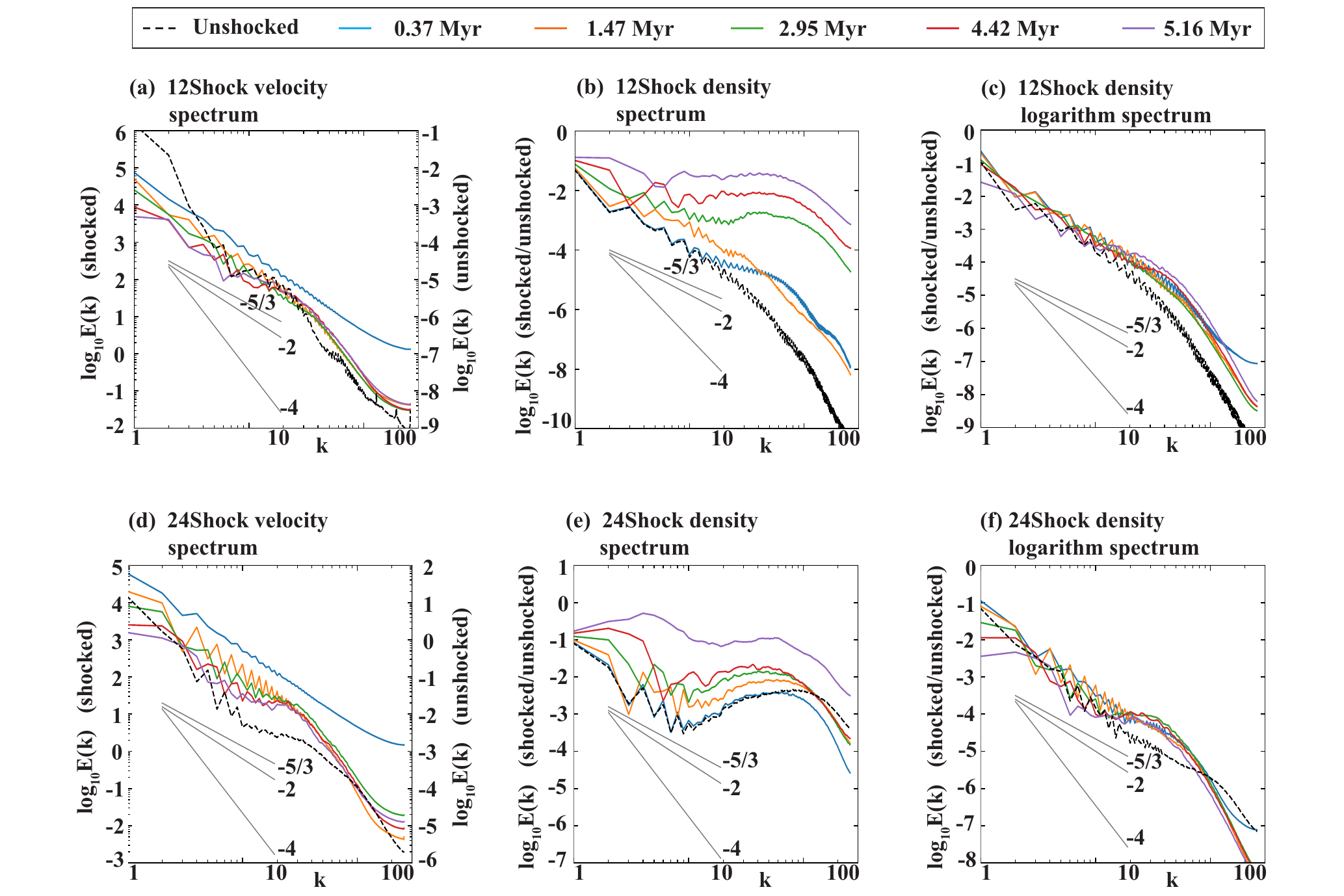}
  \caption{Power spectra of velocity (left panels), density (middle panels) and logarithmic density (right panels) for models \textit{12Shock} (top panels a\,--\,c) and \textit{24Shock} (bottom panels d\,--\,f). The colours correspond to different snapshots in time which is given as the time elapsed since shock introduction. The Kolmogorov -5/3 slope, Burgers -2 slope and a -4 slope is shown for comparison. The black dashed line is the unshocked spectra immediately prior to shock introduction. The shock adds substantial power on all scales in the velocity spectra and to include them in the same plot, the unshocked scale is shown on the right \textit{y}-axis. The scales for density are the same for both the shocked and unshocked spectra.}
\end{figure*}

\subsection{Turbulence power spectrum}

In this section we present the power spectra of velocity and density using the same IDL script that was used in WFP19, WPF20. We obtain the spectra from the square of the Fourier coefficients, calculated over a subsection of the computational domain using the 3D Fourier transform. The squares of the coefficients are then binned into wave-numbers to obtain the 1D power spectra E(\textit{k}). The physical volume used for our calculations is a 3D 150\,pc box centered at the origin and the grid is uniform and fully refined. In regions where the grid was not originally fully refined the values in the cells are interpolated to the finest level. The density power spectra are calculated using the data from the volumetric mass density and the velocity spectra from the magnitude of the complete velocity vector \textit{v}\,=\,$\sqrt{v_x^2 + v_y^2 + v_z^2}$. 

One focus in this section is to identify if any insight into the dynamics can be gained from the power spectra, and in particular if there exists an inertial range with self-similar scaling where E(k)\,$\sim$\,$k^{\alpha}$ scales with a constant value of $\alpha$ across a range in wavenumber. Two values of $\alpha$ are important to identify. Sub-sonic and incompressible turbulence approaches the Kolmogorov index of $\alpha$\,=\,-5/3 \citep{kolmogorov1941local}. In this limit, both the density and velocity spectra are expected to behave in a similar fashion as the density field is not affected by fluctuations that are characteristic of compressible flows. Supersonic and compressible turbulence approaches the limit of Burgers turbulence with an $\alpha$\,=\,-2 \citep{burgers1948mathematical}. This slope is expected to emerge only in the velocity spectrum as compressibility can significantly affect the density structures due to the presence of shocks. This can in turn flatten the slope of the density spectrum, as more power is induced on the small scales \citep[e.g.][]{kritsuk2007statistics}. Flattening of the slope in the density spectrum is also expected when self-gravity starts to be effective \citep[e.g.][]{federrath2013star}. 

We would like to stress that the power spectrum is an incredibly sensitive diagnostic, can change depending on the method used to calculate it, and provides many challenges for comparison between simulations and observations \citep{lazarian2009obtaining}. For example, different results emerge from simulations when considering parallel or perpendicular components of the transformed wavevector (although most relevant to MHD) \citep{goldreich1995toward}, when the velocity vector is projected or smoothed into two components on a plane \citep{medina2014turbulence}, or when solenoidal or compressive components from a Helmholtz decomposition of the velocity field are recovered \citep[e.g.][]{federrath2013universality,padoan2016supernova}. Additionally, flat density spectra obtained from simulations have been shown by \citet[][]{kowal2007density} to exhibit Kolmogorov-like behaviour when the logarithm of the density is considered instead. They note that use of the logarithm stops extreme values of density from distorting the spectra, and is effective for regions with high density contrasts.

In Fig.\,11 we show snapshots of velocity, density and logarithmic density power spectra for both cases \textit{12Shock} (top panels a\,--\,c) and \textit{24Shock} (bottom panels d\,--\,f). The left panels (a) and (d) show the velocity spectra, middle panels (b) and (e) show the density spectra and the right panels (c) and (f) show the density logarithm spectra. Lines with slopes $\alpha$\,=\,-5/3, -2 and -4 are also shown. 

The shock injects a substantial amount of power in the velocity spectra. It is striking that at the initial stages of the interaction they appear to display a large inertial range spanning all scales of the computational domain and the slope follows a power law somewhat close to the Burgers $\alpha$\,=\,-2. This is peculiar as it is indicative of a supersonic, compressible turbulent energy cascade which is unlikely to have had a chance to develop on these time-scales, especially since the initial shock is weak and the post-shock flow is subsonic. Tests show the shock has injected energy on all scales and monitoring snapshots immediately after the shock is introduced and following its evolution for 60\,kyr, the velocity power spectra were the same as those at 0.37\,Myrs shown in Fig.\,11(a,d) in terms of overall power and slope. Tests isolating only the shock without a cloud present on the grid also reveal such spectra. 

At 1.47\,Myrs, the velocity spectra have lost power on all scales, with most significant losses occurring after \textit{k}\,$\gtrsim$\,30 (\textit{l}\,$\lesssim$\,5\,pc) where the spectra appear to have $\alpha$\,$\approx$\,-4. In the \textit{12Shock} case, this implies that the onset of instabilities and radiative cooling is enough to obstruct energy transfer to scales smaller than 5\,pc and the break could be an indicator of the length-scale of the instabilities. These instabilities subsequently develop into cold dense clumps. Material flows onto these clumps which have a size scale on the order of 5\,pc and decelerates as it crosses the phase boundary, converting kinetic energy to gravitational energy. Hence the 5\,pc scale could be considered the dissipative limit for these models.  In the \textit{24Shock} case this could be an indicator of the length scale of the thermal instability formed structure, as there is plenty of already established structure on the grid which decelerate the incident flow. Thus as the models evolve, little change is seen in the velocity spectra at \textit{k}\,$>$\,30. We note that if there were artificial resolution effects present, we would expect to see a characteristic upturn at large wavenumber that would be indicative of resolution issues, such as we have observed in previous low-resolution aspects of suites of MHD turbulence simulations \citep{wareing2009forward,wareing2010hollerbach}. As we do not see this here, we argue that the spectra are well-resolved and the spectral break with steepening convergence at smaller scales (larger \textit{k}) is representative of physical processes in our simulations.

Most of the changes in the velocity spectra are seen at smaller \textit{k} as the slope appears to approach somewhat closer to a Kolmogorov index of -5/3. By 5.16\,Myrs, the \textit{24Shock} case looks to have an inertial range with $\alpha$\,=\,-5/3 lying approximately between 6\,$<$\,\textit{k}\,$<$\,30 (24\,pc\,$>$\,\textit{l}\,$>$\,5\,pc) and the \textit{12Shock} case appears to have a larger inertial range from 1\,$<$\,\textit{k}\,$<$\,30 (150\,pc\,$>$\,\textit{l}\,$>$\,5\,pc). It does not look as though this happens due to power increasing at smaller scales, indicative of an energy cascade, but instead due to power diminishing on the larger scales. This is most likely due to the cloud shrinking in overall size as it is compressed by the external pressure, resulting in less power on larger scales. This is also seen in the \textit{24Shock} spectra. The velocity power spectra have thus captured the overall compression of the cloud and how interactions on scales 75\,pc\,$\gtrsim$\,\textit{l}\,$\gtrsim$\,5\,pc are able to produce a turbulence-like slope with -2\,$\lesssim$\,$\alpha$\,$\lesssim$\,-5/3.

The density spectra, in both the \textit{12Shock} and \textit{24Shock} case, show drastic differences when compared to the velocity spectra. With increasing time they show significant flattening and a gain in power across all length scales. When the logarithm is considered, however, the spectra look a lot more stable. Except for the largest scales, they do not vary significantly in overall power between snapshots and do not flatten as the cloud evolves. A bump is seen around \textit{k}\,=\,20 (\textit{l}\,=\,7.5\,pc) in both cases and is most pronounced in the \textit{24Shock} case, reflecting the high density clumps on those scales. In both cases, the logarithm spectra appear to show Kolmogorov-like behaviour in an inertial range of 1\,$<$\,\textit{k}\,$<$\,20. This is in agreement with \citet{kowal2007density} who find logarithmic density spectra show this behaviour when the regular density spectra do not necessarily.

\section{Conclusion}

In this work we present a self-consistent model of the formation of a molecular cloud out of a diffuse atomic medium subject to only thermal instability (TI) and gravity. A shock is introduced at two different times and the cloud is evolved until the first instances of local gravitational collapse. During its evolution we study the relative importance of the shock, TI and gravity and note the following outcomes:

\begin{enumerate}
    \item Both shock scenarios show early and sustained evidence of local gravitational collapse, successfully demonstrating their capacity for star formation. Local collapse was not seen in the unshocked clouds as in our \textit{NoShock} case or WPFVL16, though more massive clouds do show local collapse -- see WFP19.
    \item Introducing the shock whilst the cloud is atomic prevents the development of the thermal instability. Gas is shocked into the thermally unstable regime, but cools directly to the cold phase.
    \item Radiative post-shock material cools to form a dense shell which fragments due to dynamical instabilities. Some fragments eventually collapse due to gravity.
    \item While the transmitted shock can potentially trigger the thermal instability, this is prevented because the post-shock gas that cools back to the unstable phase is repeatedly shocked on a cloud-crushing time-scale. A shock interacting with a much larger cloud, like that of WFP19, would be a better candidate for witnessing shock triggered thermal instability.
    \item TI formed clumps are important in determining the structure of the velocity field as external flows are directed via low-density inter-clump channels. This causes it to be turbulent-like. However, there is large-scale order to the flow as the majority of it is directed through the channels to the centre of the cloud and back upstream against the cloud drag.
    \item When structure is already present in the cloud, the shock substantially increases the probability of clump-clump interactions. These occurrences are largely responsible for the first instances of gravitational collapse.
    \item In our models the shocked molecular clouds remain starless for $\sim$\,5\,Myrs in the \textit{12Shock} case and $\sim$\,15\,Myrs in the \textit{24Shock} case, thus reflecting the possible lifetimes of molecular clouds prior to star formation.
    \item Both shocked simulations show turbulent-like velocity and logarithmic density power spectra. Power spectra of non-logarithmic density show drastic differences between snapshots and are very sensitive to the interaction. The power spectra in general, however, do not convey key information about the nature of the interaction. This was better captured by inspecting the cloud dynamics and morphology, and the mass-weighted pressure-density distribution. 
\end{enumerate}

This work is part of a series that will explore the effects of massive star feedback on their environments, where the clouds in this work will serve as initial conditions for future studies. In future work, therefore, we will include a robust star particle formation technique to study the influence of feedback on the already formed structure and compute the star formation rates and efficiencies.

\section*{Acknowledgements}

MMK acknowledges support from the Science and Technology Facilities Council (STFC) for PhD funding (STFC DTP grant ST/R504889/1). CJW, JMP SAEGF acknowledge STFC for project funding (STFC Research Grant ST/P00041X/1). The calculations herein were performed on ARC3, part of the High Performance Computing facilities at the University of Leeds, UK. Additional simulations were performed using the DiRAC Data Intensive service at Leicester, operated by the University of Leicester IT Services, which forms part of the STFC DiRAC HPC Facility (www.dirac.ac.uk). The equipment was funded by BEIS capital funding via STFC capital grants ST/K000373/1 and ST/R002363/1 and STFC DiRAC Operations grant ST/R001014/1. DiRAC is part of the National e-Infrastructure. We thank David Hughes at Leeds for the provision of IDL scripts which formed the basis of the power spectra analysis presented in this work. We thank the anonymous referee for feedback and corrections to the text. We also would like to thank S. Van Loo for constructive discussions and help with numerics.

\section*{Data Availability}

The dataset associated with this article is available in the Research Data Leeds Repository at https://doi.org/10.5518/903.

\appendix
\section{Thermal instability, thermal conduction and resolution}
\begin{figure*}\label{fig:tests}
  \includegraphics[width=0.75\textwidth]{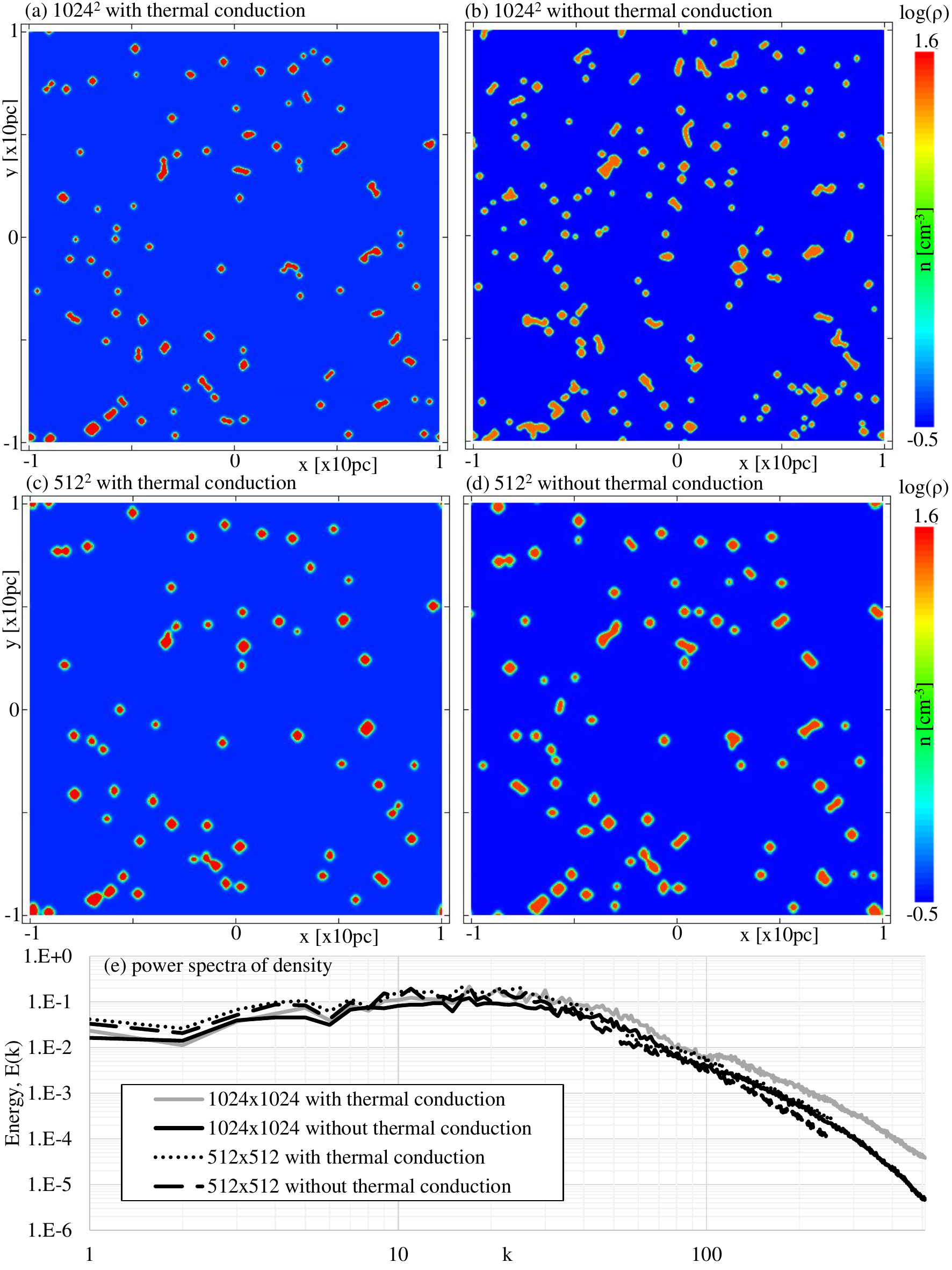}
  \caption{Results from test simulations showing left panels (a and c): simulations with thermal conduction (\textit{TC}), right panels (b and d): simulations without thermal conduction (\textit{NoTC)}. Two resolution considerations are shown with top panels (a and b): 1024$^2$ box and bottom panels (c and d): 512$^2$ box. Panel (e) shows the density power spectra for all tests.}
\end{figure*}

\citet{field1965thermal} showed that thermal conduction induces a maximum in the growth rate of the condensation mode of the thermal instability (TI) and stabilises modes whose wavelength is smaller than the Field length. If perturbations smaller than the Field length cannot be stabilised then smaller wavelengths, even down to grid scale, become unstable. In principle therefore simulations of the TI not including thermal conduction become resolution dependent and convergence becomes impossible. \citet{koyama2004field} (hereafter KI04) performed tests of the TI with and without thermal conduction. They found that in both cases, an increase in resolution increased the number of condensations on the domain. However, including thermal conduction set a limit to this when their Field length was resolved by at least 3 cells. They argue therefore that both of these conditions must be met to accurately simulate the TI.

There are two things that need to be considered. Firstly, our results and tests find agreement with KI04, where increasing resolution in simulations without thermal conduction does indeed produce more clumps. What we see however, which KI04 do not comment on, is that with or without thermal conduction the majority of the mass is still contained in the most massive clumps. When this is accounted for, the final established states of the TI are not so different between models. Where we see the biggest differences is during the development stage of the TI, where an increase in resolution results in a larger variance of initial sizes and growth rates of condensations. However it is then at the stage when structures have established where we perform our calculations (or prior for the \textit{12Shock} case).

Secondly, their definition of the Field length is inaccurate, yielding an incorrect conclusion that it must be resolved by at least 3 cells. KI04 use

\begin{equation}\label{eq:KI04field}
    \lambda_\textsc{f} = \left(\frac{T\kappa}{\rho L_c}\right)^{0.5}
\end{equation}

\noindent where \textit{L}$_{c}$ is the magnitude of the cooling rate per unit mass and $\kappa$ is the thermal conductivity. This definition does not account for derivatives of the source term, and consequently is much smaller and does not go to infinity at the boundaries of the unstable region. A more modern derivation by \citet{falle2020thermal} gives a more realistic expression for the Field length as 

\begin{equation}\label{eq:Fallefield}
\lambda_\textsc{f} = 2\pi\left(\frac{T\kappa}{\rho(\rho L_\rho -  TL_\textsc{T})}\right)^{0.5},
\end{equation}

\noindent where \textit{L}$_\rho$ and \textit{L}$_\textsc{t}$ are the derivatives of the energy loss rate per unit mass, L, w.r.t. density and temperature. Note that this is only defined for isobaric instability, $\rho$\textit{L}$_\rho$ - \textit{TL}$_\textsc{T}$\,$>$\,0, and $\lambda_\textsc{f} \to \infty$ at the boundaries of the unstable region. This agrees with equation (26) in \citet{field1965thermal} and expressions in \citet{begelman1990global} and \citet{kim2008galactic}. This is the true linear stability limit, and is shown in fig.\,3 in \citet{falle2020thermal} for our energy loss rate \citep{koyama2002origin} and a thermal conductivity given by
\begin{equation}\label{eq:k}
    \kappa = 2.5 \times 10^3T^{0.5}.
\end{equation}
\noindent\citep{parker1953instability}. Their fig. 3 also constrasts this against equation (\ref{eq:KI04field}) however we note that there is an error in the figure: equation (\ref{eq:KI04field}) has been multiplied by 2$\pi$ and so it is even lower than what is shown. We note an important consequences of this: the simulations of KI04 are actually obeying the analytical expressions derived by \citet{falle2020thermal}, and since the Field length is approximately 10\,$\times$ larger, their converged result occurring at $\Delta$\textit{x}\,$<$\,0.39$\lambda_{\textsc{f}}$ is out by a factor of 10, meaning that where they are really seeing convergence is approximately at $\Delta$\textit{x}\,$<$\,0.039$\lambda_{\textsc{f}}$, or at 20\,--\,30 cells per Field length.

We nevertheless satisfy this constraint and show a comparison in Fig.\,A1 of test simulations with (model \textit{TC}) and without thermal conduction (\textit{NoTC}). These calculations were performed on a 10\,$\times$\,10\,pc 2D periodic box seeded with $\pm$\,10 per cent random density perturbations on the grid scale around a density of \textit{n}\,=\,1.1\,cm$^{-3}$, effectively designed to represent a slice through a smaller region of the larger cloud presented as the initial condition in Section 2. A 512$^2$ and 1024$^2$ grid was used which set the resolution at $\Delta$\textit{x}\,$\approx$\,0.039\,pc and 0.0195\,pc. At the unperturbed initial density, \textit{n}\,=\,1.1, equation (\ref{eq:Fallefield}) gives $\lambda_\textsc{f}$\,=\,0.594\,pc. The Field length is therefore resolved by approximately 14 and 29 cells. Note that equation (\ref{eq:KI04field}) gives $\lambda_\textsc{f}$\,=\,0.0587\,pc, so according to this we have approximately 1.5 and 3 cells per Field length. These results are much the same as the two dimensional hydrodynamic calculations in WPFVL16 ($\Delta$\textit{x}\,=\,0.156\,pc) and our \textit{NoShock} simulations ($\Delta$\textit{x}\,=\,0.29\,pc), which have much larger grid spacing.

From inspection, it appears that the highest resolution \textit{NoTC} simulation has the most clumps and both \textit{TC} simulations experience a reduction in the amount of small-scale structure. This would indeed be consistent with KI04, however this does not appear to be reflected in the spectra. Oddly we see in Fig.\,A1(e) that the \textit{TC} models contain more power on small scales (large \textit{k}), even though the action of thermal conduction is claimed to reduce it. The claim that thermal conduction creates a reduction in small-scale structures therefore needs further testing. Nevertheless, this is not too important for our simulations, as the spectra converge on larger scales (smaller \textit{k}) where most of the power is contained, and these larger scale structures are more influential in the interaction. This agrees with the results of other authors. For example \citet{hennebelle2007structure} found that the discrepancy between their derived spectra is not very large and mostly witnessed on small scales. \cite{inoue2015thermal} noted that properties of the thermally bistable medium converged on large scales regardless if thermal conduction was included or not. Most recently, \citet{Wareing2020} performed similar tests as presented here but this time with magnetic fields. Their results, like ours, demonstrate convergence of their spectra. Note that their models with thermal conduction also contained more power on small scales than models without. However the differences between them are much smaller due to the magnetic field restricting motion to 1D thus increasing the coalescence of small clumps. Coalescence was seen in KI04 and motivated their argument, however as it is possible to achieve this with different physics and on large scales the difference is minimal, it demonstrates that thermal conduction is not a necessary ingredient for large scale TI simulations like ours, even less so for MHD.




\bibliographystyle{mnras}




\label{lastpage}
\end{document}